\documentclass[preprint,12pt]{elsarticle}

\newtheorem{prop}{Proposition}

\usepackage{amssymb}

\journal{Annals of Physics}

\begin{document}

\begin{frontmatter}

\title{Properties of a potential energy matrix in oscillator basis}

\author[BITP]{Yu. A. Lashko\corref{correspondingauthor}}
\cortext[correspondingauthor]{Yu. A. Lashko}
\ead{ylashko@gmail.com}

\author[BITP]{V. S. Vasilevsky}

\author[BITP]{G. F. Filippov}

%
\address[BITP]{Bogolyubov Institute for Theoretical Physics,\\
Metrolohichna str., 14b, Kiev, 03143, Ukraine}

\begin{abstract}
Matrix elements of potential energy are examined in detail. We consider a
model problem - a particle in a central potential. The most popular forms of
central potential are taken up, namely, square-well potential, Gaussian, Yukawa
and exponential potentials. We study eigenvalues and eigenfunctions of the
potential energy matrix constructed with oscillator functions. It is demonstrated
that eigenvalues coincide with the
potential energy in coordinate space at some specific discrete points. We
establish approximate values for these points. It is also shown that the
eigenfunctions of the potential energy matrix are the expansion coefficients of the
spherical Bessel functions in a harmonic oscillator basis. We also demonstrate
a close relation between the
separable approximation and $L^{2}$ basis (J-matrix) method for the quantum theory of scattering.
\end{abstract}


\end{frontmatter}


\section{Introduction}

Last decades, the methods which involve square-integrable bases for solving
scattering problems become popular. In atomic physics such a method is
called the J-matrix method \cite{kn:Heller1, kn:Yamani} and in nuclear physics it is known as the
algebraic version of the resonating group method \cite{kn:Fil_Okhr, kn:Fil81}.  There are two most
popular sets of basis functions which are used to tackle scattering
problems. One of them is the basis of oscillator functions, the other is the Slater
(Laguerre) basis. The key element of the J-matrix method is that the
asymptotic or reference Hamiltonian  can be represented by a tridiagonal
matrix (the Jacobi matrix) in the discrete representation (\cite{kn:Heller1, kn:Yamani}). The asymptotic Hamiltonian could be treated
analytically and it provides us with two linearly independent solutions,
much like the coordinate representation. These are the regular and irregular solutions or incoming and outgoing waves. Thus there is a full
correspondence and consistency with the scattering theory in the coordinate
representation.
Basic ideas of the J-matrix method and its applications to physical problems in different branches of
the quantum physics are outlined in a collection of papers Ref. \cite{J-matrix2008B}.

Discretization is widely used in many-body quantum physics. There are
different types of discretization. For example, it is possible to solve linear differential,
integral and integro-differential equations by involving a set of discrete
points and reducing these equations to a system of linear algebraic
equations. Alternatively, one can expand the wave function or its part (for
example, internal part) over a set of basis functions selected by
physical or computational reasons. In this case, one also arrives to a set of
homogeneous or nonhomogeneous equations. If boundary conditions have to be
implemented, this set of equations becomes nonhomogeneous. Implementation of
boundary condition can be made in the discrete representation (the J-matrix
method) or in the coordinate representation (the R-matrix method \cite{1958RvMP...30..257L}).

The essence of the J-matrix method is that it is\ the matrix form of quantum
mechanics with correct boundary conditions both for bound states and for
scattering states. Within this method the well-known boundary conditions \
in coordinate space were transformed to the discrete space of basis
functions.  Heller and Yamani \cite{kn:Heller1}, Yamani and Fishman \cite%
{kn:Yamani} demonstrated that the oscillator basis yields the tridiagonal or
Jacobi form of a reference Hamiltonian for neutral particles, while the
Laguerre basis produces the tridiagonal form for the charged particles. The
Laguerre basis has been intensively employed for solving three-body problems
in nuclear (\cite{1997PhRvC..55.1080P, 2005PhRvC..72d4003D}) and
atomic physics (\cite{2011PPN....42..683P, 2009JPhB...42d4003K}).

The simple
relation between expansion coefficients and wave function in coordinate
space has been established within the algebraic version of the resonating 
group method. It was shown in Refs. \cite{kn:Fil_Okhr, kn:Fil81} 
 that the expansion coefficients are proportional to the wave function at
discrete points of coordinate space. In Ref. \cite{OkhrimenkoC}, this relation helped 
to formulate a simple but reliable method of investigating 
reactions with charged particles by using the oscillator basis. In Ref. %
\cite{1990JPhG...16.1241M} the boundary conditions for three particles in
continuum were formulated, which allowed the application of the algebraic version of the
resonating group method for studying loosely bound states of nuclei with
a large excess of neutrons or protons \cite{2004AnPhy.312..284L}, 
resonance states and reactions in three-cluster continuum of light nuclei 
\cite{2010PPN....41..716N}.

It was also shown that the J-matrix method is equivalent to the R-matrix
method \cite{1958RvMP...30..257L}. It was demonstrated that the J-matrix method
provides a self-consistent realization of the R-matrix method.  This
relationship is discussed in the Foreword of the book \cite{J-matrix2008B} where
one can find necessary references. Being applied to the same system, the
J-matrix method and R-matrix method give a very close or identical (with the
precision of numerical procedures involved) results. Detailed historical
review, the explanation of the main ideas and modern development of the
R-matrix method are presented in Ref. \cite{2010RPPh...73c6301D}. 

Baye and coworkers developed the microscopic R-matrix method \cite%
{1977NuPhA.291..230B} which is based on the generator coordinate method. The
discrete form of the integral Hill-Wheeler equation is used to determine a
wave function in the internal region. The eigenvalues and eigenfunctions of
this equation are employed to find a total wave function and the scattering
parameters (elements of the R- or S-matrices) in accordance with the
R-matrix methodology. This method has been successfully used to study
dynamics of many-cluster systems \cite{2012LNP...848....1D}.

It is worthwhile noticing another discretization scheme which is
supplementary to the above mentioned schemes and is known as the Lagrange-mesh
method. The review of the method is given in Ref. \cite{2015PhR...565....1B}. 
Within this method the Schr\"{o}dinger equation is presented in a grid of
mesh-points generated by the Gauss-quadrature approximation. The Gauss
quadrature relates on the properties of the Lagrange polynomials and
provides the efficient and reliable solutions of the Schr\"{o}dinger
 equation. As will be seen below, the Lagrange-mesh method  overlaps
with the J-matrix method as they both employ the polynomials to expand the wave
function to be determined. 

The main differences between Lagrange-mesh method
and our realization of J-matrix method are the following (i) we do not use an approximate
expression for the kinetic energy operator and (ii) we obtain a different
approximate form of matrix elements of the potential energy operator. The
latter will be demonstrated in analytical and numerical forms.

It is also important to mention Refs. \cite{ISMAIL2011379, 
2012SIGMA...8..061I, Ismail2012..AA..327} where a rigorous analysis
has been performed to study properties of the tridiagonal equations. This
analysis gave a mathematical justification of the J-matrix method by showing
the coincidence of not only the spectra of the original Hamiltonian and the tridiagonal
matrix but also their spectral measures.
In these publications a tridiagonalizing scheme is extended for a more
general case of differential, difference and q-difference operators using
orthogonal polynomials.

Despite numerous publications on the essence of the J-matrix method and its
applications for solving model and real quantum mechanical problems, we
believe that there is a gap in studying properties of the potential energy
matrix. This matrix is a key ingredient of the J-matrix method and it determines
dynamic properties of a system under consideration. Thus our aim is to study
the main properties of the potential energy matrix in order to fill in this
gap. In the present paper, we deal with a model problem. We consider a
particle in a field of a central-symmetric potential in three-dimensional
space. In what follows we demonstrate that the model problem reveals some
very interesting properties of the potential energy matrix, which can be a
help in understanding the properties of the matrix for real nuclear systems.
In particular, the results obtained in this paper could open an effective 
and pictorial  way for studying the effect of the Pauli principle on the 
potential energy of cluster systems.

The present paper is the first one in a series of papers which are devoted to studying 
properties of  potential energy matrices in different many-body systems. 
In these papers we are going to study (i) model two-body problems (the present paper), 
(ii) two-cluster systems and (iii) three-cluster systems. As will be shown below, 
one of the important outcomes of the present paper %
 is that diagonalization of the matrix of the potential energy operator suggests a
natural way of obtaining the local form for the nonlocal operator.

The paper is organized as follows. In Sec. \ref{Sec:Method} we introduce
tools for investigating properties of the potential energy matrix and all
the main ingredients of the present calculations. In this Section we also
formulate propositions which predict the structure of eigenvalues and
eigenfunctions of the matrix. In Sec. \ref{Sec:Results} we discuss results
of numerical calculations for four potentials and verify the predictions
made in the previous Section. We give our final remarks and summarize the most
important conclusions in Sec. \ref{Sec:Conclusions}.

\section{Method \label{Sec:Method}}

\subsection{Potential energy}

In this Section, we consider an operator of potential energy and its matrix
elements between oscillator functions. As was pointed out above, we deal
with a local (in coordinate space) potential  $\widehat{V}\left(  r\right)  $.
In momentum space this potential has a nonlocal form:%
\begin{equation}
\widehat{V}_{L}\left( p,\widetilde{p}\right) =\frac{2}{\pi }\int_{0}^{\infty
}drr^{2}j_{L}\left(  pr\right)  \widehat{V}\left(  r\right)  j_{L}\left(
\widetilde{p}r\right) ,  \label{eq:SP101}
\end{equation}%
where $j_{L}\left(  x\right)  $ is a spherical Bessel function (see Section 10
of the book \cite{kn:abra}). Moreover, its shape depends on the orbital
momentum $L$.

We consider matrix elements of the potential energy operator $\widehat{V}%
\left( r\right) $\ between oscillator functions $\left\vert n\right\rangle
=\Phi _{n}\left( r,b\right) $ in coordinate space or the nonlocal operator $\widehat{V}%
_{L}\left( p,\widetilde{p}\right) $ between oscillator functions $\left\vert
n\right\rangle =\Phi _{n}\left( p,b\right) $ in momentum space, where%
\begin{eqnarray}
\Phi _{nL}\left( r,b\right) &=&\left( -1\right) ^{n}N_{nL}~b^{-3/2}\rho
^{L}e^{-\frac{1}{2}\rho ^{2}}L_{n}^{L+1/2}\left( \rho ^{2}\right) ,\quad
\rho =\frac{r}{b},  \label{eq:SP102a} \\
\Phi _{nL}\left( p,b\right) &=&N_{nL}~b^{3/2}\rho ^{L}e^{-\frac{1}{2}\rho
^{2}}L_{n}^{L+1/2}\left( \rho ^{2}\right) ,\quad \rho =pb,  \label{eq:SP102b}
\end{eqnarray}%
$b$ is the oscillator length, $n$ is the number of radial quanta and normalization
constant $N_{nL}$  is defined by
the following expression%
\begin{equation}
N_{nL}=\sqrt{\frac{2\Gamma \left( n+1\right) }{\Gamma \left( n+L+3/2\right) }%
}. \nonumber
\end{equation}%
One immediately notices the similarity of the oscillator functions in the
coordinate and momentum spaces, as they have the same dependence on the
dimensionless variable $\rho $. This reflects the similarity of the 
oscillator Hamiltonian in coordinate and momentum spaces. The oscillator 
functions are known to be related by the integral transformations.
\begin{eqnarray}
\Phi _{nL}\left( p,b\right) &=&\sqrt{\frac{2}{\pi }}\int_{0}^{\infty
}drr^{2}j_{L}\left( pr\right) \Phi _{nL}\left( r,b\right) ,
\label{eq:SP102C} \\
\Phi _{nL}\left( r,b\right) &=&\sqrt{\frac{2}{\pi }}\int_{0}^{\infty
}dpp^{2}j_{L}\left( pr\right) \Phi _{nL}\left( p,b\right) .
\label{eq:SP102D}
\end{eqnarray}%
These relations can be treated in the following way. The oscillator
functions $\Phi _{nL}\left( p,b\right) $ in the momentum space are the
expansion coefficients of the spherical Bessel function in the oscillator
functions $\Phi _{nL}\left( r,b\right) $ in the coordinate space, and 
vice versa. Recall that the spherical Bessel function $j_{L}\left( pr\right)$ 
is also a wave function of free
motion of a particle  with a fixed value of the orbital momentum $L$ and 
the wave number $p$. This assertion can be formulated as
\begin{equation}
\sqrt{\frac{2}{\pi }}j_{L}\left( pr\right) =\sum_{n=0}^{\infty }\Phi
_{nL}\left( p,b\right) \Phi _{nL}\left( r,b\right) ,  \label{eq:SP112}
\end{equation}%
which reflects the peculiarity possessed by the oscillator basis only. Another
important property of the oscillator functions, which will be employed in
the present paper, is associated with the completeness relation%
\begin{equation}
\sum_{n=0}^{\infty }\Phi _{nL}\left( x,b\right) \Phi _{nL}\left( \widetilde{x%
},b\right) =\delta \left( x-\widetilde{x}\right).  \label{eq:SP113}
\end{equation}%
Such a relation is valid for any orthonormal basis of square-integrable
functions. This relation shows us that the expansion coefficients of the
delta function (or a wave function of a free motion in the momentum space $%
x=p$) in oscillator functions are also oscillator functions.

We have to
mention the last important property of oscillator functions which is useful 
in explaining our results.
An oscillator function $%
\Phi _{nL}\left( x,b\right) $ has $n$ nodes in the region $0\leq \rho \leq
R_n$, where
\begin{equation}
R_n=\sqrt{4n+2L+3} \label{eq:SP00RN}
\end{equation}
is a turning point of the classical
oscillator with the energy $$E_{n}=\frac{\hbar ^{2}}{mb^2}\left( 2n+L+\frac{3}{2%
}\right). $$
The oscillator function has a negligibly small value in the classically forbidden
region $\rho >R_n$ due to the factor $e^{-\rho ^{2}/2}$
in Eqs. (\ref{eq:SP102a}) and (\ref{eq:SP102b}). By this reason  a
set of $N$ oscillator functions can approximate a wave function only in the range $%
0\leq \rho \leq R_N$.

Note that the matrix elements of the potential energy operator calculated 
between the oscillator functions in
coordinate space are identical to those, which are calculated in momentum
space. \ The potential energy matrix%
\begin{eqnarray}
\left\langle n L\left\vert \widehat{V}\right\vert m L\right\rangle
&=&\int_{0}^{\infty }drr^{2}\Phi _{nL}\left( r,b\right) \widehat{V}\left(
r\right) \Phi _{mL}\left( r,b\right) , \label{eq:SP105A} \\
&=&\int_{0}^{\infty }dpp^{2}\int_{0}^{\infty }d\widetilde{p}\widetilde{p}%
^{2}\Phi _{nL}\left( p,b\right) \widehat{V}_{L}\left( p,\widetilde{p}\right)
\Phi _{mL}\left( \widetilde{p},b\right)  \label{eq:SP105B}
\end{eqnarray}%
approximates the exact potential $\widehat{V}\left( r\right) $ ($\widehat{V}%
\left( p,\widetilde{p}\right) $) by the following expression%
\begin{eqnarray}
\widehat{V}_{N}\left( r,\widetilde{r}\right) &=&\sum_{n,m=0}^{N-1}\Phi
_{nL}\left(  r,b\right)  \left\langle nL\left\vert \widehat{V}\right\vert
mL\right\rangle \Phi _{mL}\left( \widetilde{r},b\right) ,  \label{eq:SP141A}
\\
\widehat{V}_{N}\left( p,\widetilde{p}\right) &=&\sum_{n,m=0}^{N-1}\Phi
_{nL}\left( p,b\right) \left\langle nL\left\vert \widehat{V}\right\vert
mL\right\rangle \Phi _{mL}\left( \widetilde{p},b\right) ,  \label{eq:SP141B}
\end{eqnarray}%
which in limiting case $N\rightarrow \infty $ coincides with the original
exact form:%
\begin{eqnarray*}
\lim_{N\rightarrow \infty }\widehat{V}_{N}\left( r,\widetilde{r}\right)
&=&\delta \left( r-\widetilde{r}\right) \widehat{V}\left( r\right), \\
\lim_{N\rightarrow \infty }\widehat{V}_{N}\left( p,\widetilde{p}\right) &=&%
\widehat{V}_{L}\left( p,\widetilde{p}\right).
\end{eqnarray*}%
As we can see, the oscillator basis realizes a specific form of separable
potentials.

Suppose we constructed $N\times N$ matrix of the operator $\widehat{V}$. \
This matrix can be reduced to the diagonal form. Let us use notations $%
\lambda _{\alpha }$ for eigenvalues and $\left\{ U_{n}^{\alpha }\right\} $\
for the corresponding eigenvectors. The latter form an orthogonal matrix $%
\left\Vert U\right\Vert $. The eigenvectors define new eigenfunctions
\begin{eqnarray}
\phi _{\alpha }\left( r,b\right) &=&\sum_{n=0}^{N-1}U_{n}^{\alpha }\Phi
_{n L}\left( r,b\right) ,  \label{eq:SP103a} \\
\phi _{\alpha }\left( p,b\right) &=&\sum_{n=0}^{N-1}U_{n}^{\alpha }\Phi
_{n L}\left( p,b\right) .  \label{eq:SP103b}
\end{eqnarray}%
It should be noted that functions $\phi _{\alpha }\left( r,b\right) $ and $%
\phi _{\alpha }\left( p,b\right) $ are formally defined in the whole
coordinate or momentum spaces. Actually, they are determined in the
restricted range of the $r$ or $p$ variable. And this is due to specific
features of the oscillator functions which have been discussed above and
because of the nonuniform convergence of a series with the oscillator
functions. The maximum value of $r$ can be defined as
\begin{equation}
r_{\max }\approx b R_N \nonumber
\end{equation}%
and the maximum value of $p$  approximately equals
\begin{equation}
p_{\max }\approx R_N/b. \nonumber
\end{equation}
Furthermore, functions $\phi _{\alpha }\left( r,b\right) $ and $\phi
_{\alpha }\left( p,b\right) $ are normalized to unity, since the oscillator
functions $\Phi _{n}\left( r,b\right) $ ($\Phi _{n}\left( p,b\right) $) are
also normalized to unity and because%
\begin{equation}
\sum_{n=0}^{N-1}\left\vert U_{n}^{\alpha }\right\vert ^{2}=1. \nonumber
\end{equation}

Thus, the potential energy matrix is represented as \
\begin{equation}
\left\Vert \left\langle nL\left\vert \widehat{V}\right\vert mL\right\rangle
\right\Vert _{N}=\left\Vert U\right\Vert ^{-1}\left\Vert
\begin{array}{cccc}
\lambda _{1} &  &  &  \\
& \lambda _{2} &  &  \\
&  & \ddots & \\
&  &  & \lambda _{N}%
\end{array}%
\right\Vert \left\Vert U\right\Vert ,  \label{eq:SP104}
\end{equation}%
and the approximate potentials $\widehat{V}_{N}\left(  r,\widetilde{r}\right)
$ from Eq. (\ref{eq:SP141A}) and $\widehat{V}_{N}\left( p,\widetilde{p}%
\right) $ from Eq. (\ref{eq:SP141B}) are transformed to the following form
\begin{eqnarray}
V_{N}\left( x,\widetilde{x}\right) &=&\sum_{\alpha =1}^{N}\lambda _{\alpha
}\phi _{\alpha }\left( x,b\right) \phi _{\alpha }\left( \widetilde{x}%
,b\right) ,  \label{eq:SP142} \\
\qquad x &=&r\quad \textrm{or} \quad x=p.  \nonumber
\end{eqnarray}

In what follows we concentrate our attention on studying properties of
eigenvalues $\lambda _{\alpha }$ of the potential energy operator and its
eigenfunctions in oscillator ($\left\{ U_{n}^{\alpha }\right\} $),
coordinate ($\phi _{\alpha }\left( r,b\right) $) and momentum ($\phi
_{\alpha }\left( p,b\right) $) representations.

\subsection{The main propositions}

Is it possible to predict the behavior of eigenvalues $\lambda _{\alpha
} $ and eigenfunctions $U_{n}^{\alpha }$? The answer to this question is
positive. Now, we formulate two propositions and after that we will justify them.

\begin{prop}
The eigenvalues $\lambda _{\alpha }$ coincide with potential energy $%
\widehat{V}\left( r\right) $ at certain discrete points of the coordinate
space.
\end{prop}

\begin{prop}
The eigenfunctions $U_{n}^{\alpha }$ coincide within a factor with the
coefficients of expansion of the spherical Bessel function (or wave function
of free motion) in the oscillator basis. And thus the expansion coefficients are the
oscillator functions within a normalization factor.

\end{prop}

\subsubsection{Justification}

To justify these propositions in a very simple way, let us use Eq. (\ref%
{eq:SP101}) which relates potentials in coordinate and momentum space. By
using this relation, we represent matrix elements $\left\langle nL\left\vert
\widehat{V}\right\vert mL\right\rangle $ in the following form%
\begin{eqnarray}
\left\langle nL\left\vert \widehat{V}\right\vert mL\right\rangle
&=&\int_{0}^{\infty }dpp^{2}\int_{0}^{\infty }d\widetilde{p}\widetilde{p}%
^{2}\Phi _{nL}\left( p,b\right) \widehat{V}_{L}\left( p,\widetilde{p}\right)
\Phi _{mL}\left( \widetilde{p},b\right)  \nonumber \\
&=&\int_{0}^{\infty }drr^{2}\widehat{V}\left( r\right) \sqrt{\frac{2}{\pi }}%
\int_{0}^{\infty }dpp^{2}\Phi _{nL}\left( p,b\right) j_{L}\left( pr\right)
\nonumber \\
&\times &\sqrt{\frac{2}{\pi }}\int_{0}^{\infty }d\widetilde{p}\widetilde{p}%
^{2}j_{L}\left( \widetilde{p}r\right) \Phi _{mL}\left( \widetilde{p},b\right)
\nonumber \\
&=&\int_{0}^{\infty }drr^{2}C_{nL}\left( r,b\right) \widehat{V}\left(
r\right) C_{mL}\left( r,b\right).  \label{eq:SP143}
\end{eqnarray}%
Here $C_{nL}$ are the expansion coefficients of the spherical Bessel
function or wave function of free motion of a particle with the energy $E=%
\frac{\hbar ^{2}p^{2}}{2m}$%
\begin{equation}
\Psi _{EL}^{\left( 0\right) }=\sqrt{\frac{2}{\pi }}j_{L}\left( pr\right)
\label{eq:SP100}
\end{equation}%
in the oscillator basis%
\begin{equation}
C_{nL}=\left\langle n|\Psi _{EL}^{\left( 0\right) }\right\rangle =\sqrt{\frac{%
2}{\pi }}\int_{0}^{\infty }dpp^{2}\Phi _{nL}\left( p,b\right) j_{L}\left(
pr\right). \nonumber
\end{equation}%
These coefficients $C_{nL}$ are identical to  the oscillator functions $\Phi
_{nL}\left(  r,b\right)  $ in coordinate space. This results from the specific
properties of the oscillator functions demonstrated, in particular, in Eqs. (\ref%
{eq:SP102a}) and (\ref{eq:SP102b}).

To evaluate the integral in Eq. (\ref%
{eq:SP143}), we can use one of the well-known schemes\ of the discrete
approximation of definite integrals (see, for example, Chapter 25\ of Ref.
\cite{kn:abra} or Ref. \cite{Krylov_Integral} for more details).
\begin{eqnarray}
\left\langle nL\left\vert \widehat{V}\right\vert mL\right\rangle
&=&\int_{0}^{\infty }drr^{2}C_{nL}\left( r,b\right) \widehat{V}\left(
r\right) C_{mL}\left( r,b\right)  \label{eq:SP144} \\
&\approx &\sum_{\alpha =1}^{N}W_{\alpha }r_{\alpha }^{2}C_{nL}\left(
r_{\alpha },b\right) \widehat{V}\left( r_{\alpha }\right) C_{mL}\left(
r_{\alpha },b\right) ,  \nonumber
\end{eqnarray}%
In these discrete schemes, the discrete coordinates $r_{\alpha }$ are zeros
of some related polynomials, and the weights $W_{\alpha }$ are also related
to such polynomials. By introducing the following notation

\begin{equation}
\overline{C}_{nL}\left( r_{\alpha },b\right) =\sqrt{W_{\alpha }}r_{\alpha
}C_{nL}\left( r_{\alpha },b\right)  \label{eq:SP145}
\end{equation}%
we represent the relation (\ref{eq:SP144}) as%
\begin{equation}
\left\langle nL\left\vert \widehat{V}\right\vert mL\right\rangle \approx
\sum_{\alpha =1}^{N}\overline{C}_{nL}\left( r_{\alpha },b\right) \widehat{V}%
\left( r_{\alpha }\right) \overline{C}_{mL}\left( r_{\alpha },b\right) .
\label{eq:SP146}
\end{equation}%
Equations (\ref{eq:SP146}) and (\ref{eq:SP145}) confirm both propositions
concerning the eigenvalues and the eigenfunctions of the potential energy
matrix, but there are some questions to be answered. We need to determine 
the discrete coordinates $r_{\alpha }$ and to reveal their dependence on 
the shape of the potential considered and on the number $N$ of the basis 
functions invoked.

In this subsection, we presented some justifications of our propositions. In the
next subsection we will give proof to them.

\subsection{Proof}

For simplicity, we prove these propositions for a specific type of potentials 
that contain only even powers of coordinate $r$ in the Taylor series
\begin{equation}
\widehat{V}\left( r\right) =\sum_{\nu =0}^{\infty }\frac{V^{\left( \nu
\right) }\left( 0\right) }{\nu !}r^{2\nu },  \label{eq:SP151}
\end{equation}%
where%
\begin{equation}
V^{\left( \nu \right) }\left( 0\right) =\left. \frac{d^{\nu }}{dr^{\nu }}%
V\left( r\right) \right\vert _{r=0}. \nonumber
\end{equation}%
There are several types of potential with such properties, such as a Gaussian
 potential, P\"{o}schl-Teller potential and so on. There are several other
potentials which can be presented as an integral with Gaussian functions.
They are Coulomb, Yukawa, and exponential potentials.

Equation (\ref{eq:SP151}) suggests that matrix elements of such potentials
between the oscillator functions can be represented as%
\begin{equation}
\left\langle nL\left\vert \widehat{V}\right\vert mL\right\rangle =\sum_{\nu
=0}^{\infty }\frac{V^{\left( \nu \right) }\left( 0\right) }{\nu !}%
\left\langle nL\left\vert r^{2\nu }\right\vert mL\right\rangle .
\label{eq:SP152}
\end{equation}%
Let us denote the matrix of $r^{2}$ by%
\begin{equation}
\mathbf{R}=\left\Vert \left\langle nL\left\vert r^{2}\right\vert
mL\right\rangle \right\Vert \nonumber
\end{equation}%
By using completeness relation for oscillator functions, it easy to show that%
\begin{equation}
\left\Vert \left\langle nL\left\vert r^{2\nu }\right\vert mL\right\rangle
\right\Vert =\mathbf{R}^{\nu }.  \label{eq:SP153}
\end{equation}%
We demonstrate this relation for the operator $r^{4}$:%
\begin{equation}
\left\langle nL\left\vert r^{4}\right\vert mL\right\rangle =\left\langle
nL\left\vert r^{2}\times r^{2}\right\vert mL\right\rangle =\sum_{k=0}^{\infty
}\left\langle nL\left\vert r^{2}\right\vert kL\right\rangle \left\langle
kL\left\vert r^{2}\right\vert mL\right\rangle =\mathbf{R}^{2}.  \nonumber
\end{equation}%
One can repeat this procedure for an arbitrary power of $r^{2}$ and obtain
the relation (\ref{eq:SP153}). \ This relation allows  rewriting matrix
elements $\left\langle nL\left\vert \widehat{V}\right\vert mL\right\rangle $
of the potential energy in the form%
\begin{equation}
\left\Vert\left\langle nL\left\vert \widehat{V}\right\vert mL\right\rangle \right\Vert=\sum_{\nu
=0}^{\infty }\frac{V^{\left( \nu \right) }\left( 0\right) }{\nu !}\mathbf{R}%
^{\nu }=\mathbf{V}\left( \mathbf{R}\right)  \label{eq:SP154}
\end{equation}%
where $\mathbf{V}\left( \mathbf{R}\right) $\ is a function of the matrix $%
\mathbf{R}$. The general definition of a function of a matrix can be found in Chapter 5 of
book \cite{Ganmacher_Matrix}.
We use a very important feature of a function of the matrix described in this book:
 if two matrices $A$ and $B$ are similar and they are related
by matrix $T$%
\begin{equation}
B=T^{-1}AT, \nonumber
\end{equation}%
then functions of the matrices $f\left(  B\right)  $\ and $f\left(  A\right)  $ are
also similar%
\begin{equation}
f\left( B\right) =T^{-1}f\left( A\right) T . \nonumber
\end{equation}%
For our aims, it means that if we reduce matrix $\mathbf{R}$ to diagonal form
(or to use the spectral decomposition of the matrix)%
\begin{equation}
R=U^{-1}DU \nonumber
\end{equation}%
where $D$ is a diagonal matrix%
\begin{equation}
D=\left\Vert
\begin{array}{cccc}
r_{1}^{2} &  &  & \\
& r_{2}^{2} &  & \\
&  & r_{3}^{2} & \\
&  &  & \ddots%
\end{array}%
\right\Vert  \label{eq:SP160}
\end{equation}%
consisting of the eigenvalues $r_{\alpha }^2$ ($\alpha $=1, 2, \ldots ) of the $%
\mathbf{R}$ matrix and $U$\ is an orthogonal matrix, then
\begin{equation}
R^{\nu }=U^{-1}D^{\nu }U  \label{eq:SP161}
\end{equation}%
and
\begin{equation}
D^{\nu }=\left\Vert
\begin{array}{cccc}
r_{1}^{2 \nu } &  &  &  \\
& r_{2}^{2 \nu } &  &  \\
&  & r_{3}^{2 \nu } &  \\
&  &  & \ddots%
\end{array}%
\right\Vert .  \label{eq:SP162}
\end{equation}

Combining Eqs. (\ref{eq:SP154}), (\ref{eq:SP161}) and Eq. (\ref{eq:SP162}),
we obtain that the matrix $\mathbf{V}\left( D\right) $ is also a diagonal
matrix%
\begin{equation}
\mathbf{V}\left(  D\right)  =\left\Vert
\begin{array}{cccc}
V\left(  r_{1}\right)  &  &  & \\
& V\left(  r_{2}\right)  &  & \\
&  & V\left(  r_{3}\right)  & \\
&  &  & \ddots%
\end{array}%
\right\Vert \nonumber
\end{equation}%
consisting of potential energy $V\left( r\right) $ at some discrete points $%
r_{\alpha }$. Consequently, the matrix of potential energy can be decomposed
as
\begin{equation}
\mathbf{V}\left( \mathbf{R}\right) =U^{-1}\mathbf{V}\left( D\right) U.
\label{eq:SP164}
\end{equation}%
Thus, by reducing the potential energy matrix to the diagonal form, we
obtain eigenvalues which are equal to
\begin{equation}
\lambda _{\alpha }=V\left( r_{\alpha }\right) .  \label{eq:SP165}
\end{equation}

What do we know about eigenvalues and eigenfunctions of the matrix $\mathbf{R}$
or, in other words, of the operator $r^{2}$? The matrix $\mathbf{R}$ has a
tridiagonal form with matrix elements%
\begin{equation}
\left\langle mL\left\vert r^{2}\right\vert nL\right\rangle =b^{2}\left\{
\begin{array}{cc}
\sqrt{n\left( n+L+\frac{1}{2}\right) } & m=n-1 \\
\left( 2n+L+\frac{3}{2}\right) & m=n \\
\sqrt{\left( n+1\right) \left( n+L+\frac{3}{2}\right) } & m=n+1%
\end{array}%
\right.  \label{eq:SP166A}
\end{equation}%
This matrix is similar to the matrix of the kinetic energy operator%
\begin{equation}
\left\langle mL\left\vert \widehat{T}\right\vert nL\right\rangle =\frac{\hbar
^{2}}{2mb^{2}}\left\{
\begin{array}{cc}
-\sqrt{n\left( n+L+\frac{1}{2}\right) } & m=n-1 \\
\left( 2n+L+\frac{3}{2}\right) & m=n \\
-\sqrt{\left( n+1\right) \left( n+L+\frac{3}{2}\right) } & m=n+1%
\end{array}%
\right.  \label{eq:SP166B}
\end{equation}%

The eigenfunction of the
operator $p^{2}$ (or $\widehat{T}$ ) is the spherical Bessel function $\sqrt{%
\frac{2}{\pi }}j_{L}\left( kr\right) $ in the coordinate space and the delta
function $\delta \left( p-k\right) $ in the momentum space, while the
eigenfunction of the operators $r^{2}$ is the delta function in the
coordinate space and the spherical Bessel function in the momentum space.
Consequently, in the oscillator representation the eigenfunctions of both
operators are  coefficients of the expansion of the spherical Bessel function or
delta function in the oscillator basis. In other words, by taking into account relations (\ref%
{eq:SP112}), (\ref{eq:SP103a}) and (\ref{eq:SP103b}), we conclude that the eigenfunctions of
operators $r^{2}$ and $p^{2}$ in the discrete representation are proportional to
the oscillator functions in coordinate (\ref{eq:SP102a}) or in momentum (\ref%
{eq:SP102b}) space, correspondingly. Thus, the eigenfunctions of \ the $%
N\times N$ matrix of the operator $r^{2}$ in the discrete representation are
\begin{equation}
U_{n}^{\alpha }=\mathcal{N}_{\alpha }\Phi _{nL}\left( r_{\alpha },b\right) ,
\label{eq:SP167}
\end{equation}%
where $\mathcal{N}_{\alpha }$ is a normalization factor which can be
determined from the normalization condition%
\begin{equation}
\sum_{n=0}^{N-1}\left\vert U_{n}^{\alpha }\right\vert ^{2}=\mathcal{N}%
_{\alpha }^{2}\sum_{n=0}^{N-1}\left\vert \Phi _{nL}\left( r_{\alpha
},b\right) \right\vert ^{2}=1. \nonumber
\end{equation}
It is obvious that the eigenvalues $\lambda _{\alpha }$, the normalization
factor $\mathcal{N}_{\alpha },$ and discrete coordinates $r_{\alpha }$ depend on
the size $N$ of the matrix of the potential energy operator.

It has been shown \cite{J-matrix2008B} that the eigenvalues $r_{\alpha}^{2}$ 
and $\mathcal{E}_{\alpha}=\frac{\hbar^{2}}%
{2m}p_{\alpha}^{2}$
of the operators $r^{2}$ and $\widehat{T}$, respectively, can be obtained
from the conditions
\begin{eqnarray}
U_{N}^{\alpha } &=&\mathcal{N}_{\alpha }\Phi _{NL}\left( r_{\alpha
},b\right) =0  \label{eq:SP169A} \\
U_{N}^{\alpha } &=&\mathcal{N}_{\alpha }\Phi _{NL}\left( p_{\alpha
},b\right) =0  \label{eq:SP169B}
\end{eqnarray}%
Thus the zeros of the Laguerre polynomial $L_{N}^{L+1/2}\left( \rho \right) $
($\rho =r_{\alpha }/b$ for the operator $r^{2}$ \ and $\rho =p_{\alpha }b$
for the operator $\widehat{T}$) determine the discrete variables $r_{\alpha}$ 
and $p_{\alpha}$. To evaluate the zeros of the oscillator
function $\Phi _{NL}\left( r_{\alpha },b\right) $, we assume that $N$ is large  ($N\gg
1$) and $\rho $\ is small ($\rho \ll R_N$). In this case
we can refer to the following asymptotic form for the oscillator functions (see
Eq. (22.15.2)\ of book \cite{kn:abra} )%
\begin{equation}
\Phi _{NL}\left( r_{\alpha },b\right) \approx \left( -1\right) ^{n}\frac{2}{%
\sqrt{\pi }}\frac{1}{b^{3/2}}\sqrt{R_{N}}j_{L}\left( \rho R_{N}\right)
\label{eq:SP170}
\end{equation}%
where $R_{N}$
is the turning point for the classical harmonic oscillator discussed above (\ref{eq:SP00RN}).
Assuming that an argument of the Bessel function $\rho R_{N}$ is large ($%
\rho R_{N}\gg 1$), we use its asymptotic form
\begin{equation}
\Phi _{NL}\left( r_{\alpha },b\right) \approx \left( -1\right) ^{n}\frac{2}{%
\sqrt{\pi }}\frac{1}{b^{3/2}}\frac{1}{\rho \sqrt{R_{N}}}\sin \left( \rho
R_{N}-\frac{\pi }{2}L\right). \nonumber
\end{equation}%
From this equation we obtain
\begin{equation}
\rho R_{N}-\frac{\pi }{2}L=\alpha \pi \nonumber
\end{equation}%
or, finally,%
\begin{equation}
r_{\alpha }=\frac{\pi b\left( \alpha +\frac{1}{2}L\right) }{R_{N}}
\label{eq:SP171}
\end{equation}%
for $\alpha $=1, 2, \ldots , $N$. Equation (\ref{eq:SP171}) represents the
approximate estimates of the eigenvalues of the operator $r^{2}$.

It is important to underline, that as was stated above, the diagonalization
of $N\times N$ matrix of the operator $r^{2}$ and $p^{2}$ (or $\widehat{T}$
) reveals those eigenvalues and their eigenfunctions which obey the
"boundary condition" $U_{N}^{\alpha }=0$, i.e. the next to the last
expansion coefficient is equal to zero. This is also true for Hamiltonians
of two-body, two- and three-cluster systems. One can see some illustrations
to this statement in Refs. \cite{2015NuPhA.941..121L, 2018PhRvC..97f4605V}. 
 Such a "boundary condition" for the operator $r^{2}$ and
 $p^{2}$ is due to the tridiagonal form of these matrices. As for the two-body and
two-cluster Hamiltonians, this condition is due to the dominance of the kinetic energy
 over the potential energy.

However, this "boundary condition" is not correct for the operators $\mathbf{R%
}^{\nu }$ with $\nu >$1 and, consequently, for the potential energy matrix.
This will be demonstrated in Section \ref{Sec:Results}. We will also demonstrate that
for every potential and a given value of the oscillator length $b$, there is a
point $N_{0}>N$ in the oscillator space where the expansion coefficients have a node.
By using the explicit form (\ref{eq:SP167}) of the expansion
coefficients $U_{n}^{\alpha },$ we will easily extrapolate them to an arbitrary
value of the index $n$ of the oscillator function and find  the number of
quanta $n=N_{0}$ corresponding to such a node.

In all these relations it was implicitly assumed that we deal with large but
finite matrices.

The proof of the propositions is complete.

To calculate matrix elements and to prove the Preposition for Yukawa ($P=Y$%
), exponential ($P=E$) and the Coulomb ($P=C$) potential, which
cannot be directly expanded over the even powers of coordinate, we used the
integral transformations which relates these potentials and Gaussian
potential: 
\begin{equation}
\widehat{V}_{P}\left( r\right) =V_{0}\int_{0}^{\infty }dxg_{P}\left(
x\right) \exp \left\{ -x^{2}r^{2}\right\}  \label{eq:206}
\end{equation}%
where 
\begin{eqnarray*}
g_{Y}\left( x\right) &=&\frac{2a}{\sqrt{\pi }}\exp \left\{ -\frac{1}{%
4x^{2}a^{2}}\right\} ,~g_{E}\left( x\right) =\frac{1}{a\sqrt{\pi }}\frac{1}{%
x^{2}}\exp \left\{ -\frac{1}{4x^{2}a^{2}}\right\} \\
g_{C}\left( x\right) &=&\frac{2}{\sqrt{\pi }}
\end{eqnarray*}
Thus, such integral relations allow us to apply the above procedure to
this class of potentials. By using this integral transformations and
expanding the exponent $\exp \left\{ -x^{2}r^{2}\right\} $ we obtain the
modified version of Eq. (\ref{eq:SP152}) 
\begin{equation}
\left\langle n\left\vert \widehat{V}\right\vert m\right\rangle
=V_{0}\int_{0}^{\infty }dxg\left( x\right) \sum_{\nu =0}^{\infty }\frac{%
\left( -1\right) ^{\nu }x^{2\nu }}{\nu !}\left\langle n\left\vert r^{2\nu
}\right\vert m\right\rangle  \label{eq:SP152M}
\end{equation}%
where $g\left( x\right) $ is one of three weight functions from Eq. (\ref%
{eq:206}).

We decided to restrict ourselves with the proof the Prepositions for such
type of potentials $\widehat{V}\left( r\right) $ which can directly 
or through an integral transformation be expanded over the even powers of coordinate $r$.
There are two reasons for such decision. First, the used proof covers a
large variety of model potentials. We assume and will later confirm that the
conclusions made about eigenvalues and eigenfunctions of these potentials
are valid for a general form of two-body potentials. Second, the proof of the
Propositions for a general form of the potentials requires very lengthy explanation
which, as we believe, will yield the same results.


\subsection{Utilization of separable form}

One can see that oscillator basis proposes a new separable form of the
initial potential. It is well-known that a separable representation
simplifies finding of a solution of the Schr\"{o}dinger equations of two-
and many-particle systems. Different forms of separable representation and
solutions of the Schr\"{o}dinger equation with such type of potentials are
thoroughly discussed in Refs.\ \cite{kn:Newton, bookBelyaevV90E,
Zubarev_PPN76}.

With the separable form of potential (\ref{eq:SP141A}) or (\ref{eq:SP141B}),
the Schr\"{o}dinger equation for a wave function of bound or scattering states
with the energy $E$ appears as
\begin{equation}
\left( T-E\right) \Psi _{EL}+\sum_{n,m=0}^{N-1}\Phi _{nL}\left( x,b\right)
\left\langle nL\left\vert \widehat{V}\right\vert mL\right\rangle \left\langle
m|\Psi _{EL}\right\rangle =0,  \label{eq:SP201}
\end{equation}%
which has a formal solution%
\begin{equation}
\Psi _{EL}=-G_{0}\left( E\right) \sum_{n,m=0}^{N-1}\Phi _{nL}\left(
x,b\right) \left\langle nL\left\vert \widehat{V}\right\vert mL\right\rangle
\left\langle m|\Psi _{EL}\right\rangle  \label{eq:SP202A}
\end{equation}%
for bound states and
\begin{equation}
\Psi _{EL}=\Psi _{EL}^{\left( 0\right) }-G_{0}\left( E\right)
\sum_{n,m=0}^{N-1}\Phi _{nL}\left( x,b\right) \left\langle nL\left\vert
\widehat{V}\right\vert mL\right\rangle \left\langle m|\Psi _{EL}\right\rangle
\label{eq:SP202B}
\end{equation}%
for scattering states. These wave functions can be written in the coordinate
($x=r$) and momentum ($x=p$) spaces. Here $G_{0}\left( E\right) $ is the
Green function%
\begin{equation}
G_{0}\left( E\right) =\left( \widehat{T}-E\right) ^{-1} \nonumber
\end{equation}%
and wave function $\Psi _{EL}^{\left( 0\right) }$ is a regular solution of
the equation%
\begin{equation}
\left( \widehat{T}-E\right) \Psi _{EL}^{\left( 0\right) }=0. \nonumber
\end{equation}%
The quantity $\left\langle m|\Psi _{EL}\right\rangle $ can be considered as
a projection of the wave function $\Psi _{EL}$ on basis state $\left\vert
m\right\rangle =\Phi _{mL}$, or as the expansion coefficients of the wave
function $\Psi _{EL}$ in basis states $\left\{ \Phi _{mL}\right\} $.

By multiplying from the left Eq. (\ref{eq:SP202B}) by the oscillator
function $\Phi _{\widetilde{m}L}\left( r,b\right) $ and integrating over
coordinate $r$, we obtain the set of equations for the expansion
coefficients $\left\{ \left\langle m|\Psi _{EL}\right\rangle \right\} $%
\begin{equation}
\left\langle \widetilde{m}|\Psi _{EL}\right\rangle =\left\langle \widetilde{m%
}|\Psi _{EL}^{\left( 0\right) }\right\rangle -\sum_{n,m=0}^{N-1}\left\langle
\widetilde{m}L\left\vert G_{0}\left( E\right) \right\vert nL\right\rangle
\left\langle nL\left\vert \widehat{V}\right\vert mL\right\rangle \left\langle
m|\Psi _{EL}\right\rangle . \nonumber
\end{equation}%
Solution of this equation is%
\begin{equation}
\left\langle m|\Psi _{EL}\right\rangle =\sum_{\widetilde{m}=0}^{N-1}M_{m,%
\widetilde{m}}^{-1}\left\langle \widetilde{m}|\Psi _{EL}^{\left( 0\right)
}\right\rangle , \nonumber
\end{equation}%
where $\left\Vert M_{m,\widetilde{m}}^{-1}\right\Vert $ is an inverse matrix
to the matrix
\begin{equation}
M_{\widetilde{m},m}=\delta _{\widetilde{m},m}+\sum_{n=0}^{N-1}\left\langle
\widetilde{m}L\left\vert G_{0}\left( E\right) \right\vert nL\right\rangle
\left\langle nL\left\vert \widehat{V}\right\vert mL\right\rangle . \nonumber
\end{equation}%
Thus, the scattering wave function is%
\begin{equation}
\Psi _{EL}=\Psi _{EL}^{\left( 0\right) }-G_{0}\left( E\right)
\sum_{n,m=0}^{N-1}\sum_{\widetilde{m}=0}^{N-1}\Phi _{nL}\left( x,b\right)
\left\langle nL\left\vert \widehat{V}\right\vert mL\right\rangle M_{m,%
\widetilde{m}}^{-1}\left\langle \widetilde{m}|\Psi _{EL}^{\left( 0\right)
}\right\rangle .  \label{eq:SP301}
\end{equation}%
By using the standard definition (see, for example, Chapter 4 of book \cite%
{BazBookE})%
\begin{equation}
t\left( p,k\right) =\left\langle \Psi _{E_{p}L}^{\left( 0\right) }\left\vert
\widehat{V}\right\vert \Psi _{EL}\right\rangle ,\quad E_{p}=\frac{\hbar
^{2}p^{2}}{2m}, \nonumber
\end{equation}%
the half-off shell T-matrix can be easily determined%
\begin{eqnarray}
t\left( p,k\right) &=&\sum_{n,m=0}^{N-1}\Phi _{nL}\left( p,b\right)
\left\langle nL\left\vert \widehat{V}\right\vert mL\right\rangle \left\langle
m|\Psi _{EL}\right\rangle  \label{eq:SP302} \\
&=&\sum_{n,m=0}^{N-1}\sum_{\widetilde{m}=0}^{N-1}\Phi _{nL}\left( p,b\right)
\left\langle nL\left\vert \widehat{V}\right\vert mL\right\rangle M_{m,%
\widetilde{m}}^{-1}\left\langle \widetilde{m}|\Psi _{EL}^{\left( 0\right)
}\right\rangle .  \nonumber
\end{eqnarray}
Similar form of the wave function and T-matrix was obtained in Ref.
\cite{kn:Heller1}.

If we take another separable form of the potential from Eq. (\ref{eq:SP142}%
), then we will get%
\begin{equation}
\Psi_{EL}=\Psi_{EL}^{\left( 0\right) }-G_{0}\left( E\right) \sum
_{\alpha=1}^{N}\lambda_{\alpha}\phi_{\alpha}\left( x,b\right) \left\langle
\phi_{\alpha}|\Psi_{EL}\right\rangle \nonumber
\end{equation}%
followed by%
\begin{equation}
\left\langle \phi_{\nu}|\Psi_{EL}\right\rangle =\left\langle
\phi_{\nu}|\Psi_{EL}^{\left( 0\right) }\right\rangle
-\sum_{\alpha=1}^{N}\left\langle \phi_{\nu}\left\vert G_{0}\left( E\right)
\right\vert \phi_{\alpha }\right\rangle \lambda_{\alpha}\left\langle
\phi_{\alpha}|\Psi_{EL}\right\rangle . \nonumber
\end{equation}%
It is easy to see that
\begin{equation}
\left\langle \phi _{\alpha }|\Psi _{EL}\right\rangle =\sum_{\nu
=1}^{N}K_{\alpha \nu }^{-1}\left\langle \phi _{\nu }|\Psi _{EL}^{\left(
0\right) }\right\rangle , \nonumber
\end{equation}%
where%
\begin{equation}
K_{\nu \alpha }=\delta _{\nu ,\alpha }+\left\langle \phi _{\nu }\left\vert
G_{0}\left( E\right) \right\vert \phi _{\alpha }\right\rangle \lambda
_{\alpha }. \nonumber
\end{equation}%
Thus, the wave function for scattering state is%
\begin{equation}
\Psi _{EL}=\Psi _{EL}^{\left( 0\right) }-G_{0}\left( E\right) \sum_{\alpha
=1}^{N}\sum_{\nu =1}^{N}\lambda _{\alpha }\phi _{\alpha }\left( x,b\right)
K_{\alpha \nu }^{-1}\left\langle \phi _{\nu }|\Psi _{EL}^{\left( 0\right)
}\right\rangle  \label{eq:SP303}
\end{equation}%
and the T-matrix equals%
\begin{eqnarray}
t\left( p,k\right) &=&\sum_{\alpha =1}^{N}\lambda _{\alpha }\phi _{\alpha
}\left( p,b\right) \left\langle \phi _{\alpha }|\Psi _{EL}\right\rangle
\label{eq:SP304} \\
&=&\sum_{\alpha =1}^{N}\sum_{\nu =1}^{N}\lambda _{\alpha }\phi _{\alpha
}\left( p,b\right) K_{\alpha \nu }^{-1}\left\langle \phi _{\nu }|\Psi
_{EL}^{\left( 0\right) }\right\rangle ,  \nonumber
\end{eqnarray}%
where
\begin{eqnarray*}
\left\langle \phi _{\nu }|\Psi _{EL}^{\left( 0\right) }\right\rangle &
=& \sum_{n=0}^{N-1}U_{n}^{\nu }\left\langle n|\Psi _{EL}^{\left( 0\right)
}\right\rangle , \\
\left\langle \phi _{\nu }\left\vert G_{0}\left( E\right) \right\vert \phi
_{\alpha }\right\rangle & =& \sum_{n,m=0}^{N-1}U_{n}^{\nu }\left\langle
nL\left\vert G_{0}\left( E\right) \right\vert mL\right\rangle U_{m}^{\alpha }.
\end{eqnarray*}%
Eq. (\ref{eq:SP302})\ and (\ref{eq:SP304}) present the half-off shell
T-matrix for the energy $E=\frac{\hbar ^{2}k^{2}}{2m}$.

Formulae (\ref{eq:SP301})--(\ref%
{eq:SP304}) suggest that to find the wave functions and the T-matrix for the
scattering states one needs to calculate (i) the expansion coefficients for
the free-motion wave function $\left\langle k|\Psi _{EL}^{\left( 0\right)
}\right\rangle $ ($\left\langle \phi _{\nu }|\Psi _{EL}^{\left( 0\right)
}\right\rangle $) and (ii) matrix elements of the Green function $%
G_{0}\left( E\right) $ between basis functions $\left\{ \Phi _{nL}\right\} $(%
$\left\{ \phi _{\alpha }\right\} $). Then, one obtains the exact form of the
wave functions and T-matrix using matrix manipulations. The explicit form
of matrix elements $\left\langle n\left\vert G_{0}\left( E\right)
\right\vert m\right\rangle $ and recurrence relations they satisfy one can
find in Ref. \cite{Heller1975}.

Moreover, Eqs. (\ref{eq:SP301})--(\ref%
{eq:SP304}) can be considered as an alternative way for determination of the
wave functions and scattering parameters. Formally these equations do not
require the asymptotic Hamiltonian to be the tridiagonal Jacobi matrix. Thus
these equations can be applied to the problem when the asymptotic
Hamiltonian includes the Coulomb interactions, provided that the expansion
coefficients of the regular Coulomb wave function and the corresponding
Green function can be calculated for such asymptotic Hamiltonian.

It is worthwhile noticing that the separable form of the potential
energy matrix (\ref{eq:SP142}) is known as the Bubnov-Galerkin
separabilization scheme \cite{bookBelyaevV90E, Zubarev_PPN76} and
can be also presented in the following form%
\[
\int dxx^{2}\int d\widetilde{x}\widetilde{x}^{2}\phi_{\alpha}\left(  x\right)
\widehat{V}_{N}\left(  x,\widetilde{x}\right)  \phi_{\beta}\left(
\widetilde{x}\right)  =\lambda_{\alpha}\delta_{\alpha\beta}.
\]
It means that to realize this scheme one needs to find a set of
functions $\left\{  \phi_{\alpha}\right\}  $ which are orthogonal
with  a potential being the weight function. Diagonalization of the potential energy matrix
calculated between the oscillator functions produces such a set of orthogonal
functions.  And this set of functions is available in the oscillator, coordinate
and momentum representations.

The correspondence between the traditional form of the J-matrix method and
the separable representation will be studied elsewhere in more detail.

\section{Results and discussion \label{Sec:Results}}

In this paper we do not dwell on the problem how to calculate matrix elements
of a potential between the oscillator functions. We just point out that we
make use of recurrence relations as one of the reliable ways for
construction matrices $\left\Vert \left\langle nL\left\vert \widehat
{V}\right\vert mL\right\rangle \right\Vert _{N}$ for an arbitrary value of N. The
main ideas of the recurrence relation method have been explained in Refs.
\cite{1994AmJPh..62..362A, kn:VA_PR}. Matrix elements obtained
with recurrence relations have been checked in Refs.
\cite{2004nucl.th..12085B, J-matrix2008.117, 
2015UkrJPh..60.297V}. Having calculated spectrum of bound states of a
model problem and phase shifts, we verified that the matrix elements were
correctly constructed. Both energies of bound state and phase shifts were
compared with those obtained by some alternative methods of calculation and
demonstrated full consistency.

We selected four two-body potentials to study properties of their matrix
elements. These potentials are very often used in atomic, molecular and
nuclear physics to model real physical processes. They are Gaussian,
exponential, Yukawa and square-well potentials determined by the relations%
\begin{eqnarray*}
\widehat{V}\left(  r\right)    & =V_{0}\exp\left\{  -z^{2}\right\}  ,\\
\widehat{V}\left(  r\right)    & =V_{0}\exp\left\{  -z\right\}  ,\\
\widehat{V}\left(  r\right)    & =V_{0}\exp\left\{  -z\right\}  /z,\\
\widehat{V}\left(  r\right)    & =\left\{
\begin{array}
[c]{cc}%
V_{0} & r\leq a\\
0 & r\geq a
\end{array}
\right.  ,
\end{eqnarray*}
Where $V_{0}$ is the depth, $a$ is the range of the potential, and $z=r/a$.

We used the same parameters of the potentials as in Ref. \cite{2015UkrJPh..60.297V}. 
Namely, the depth of all the
potentials was chosen to be $V_{0}=$-85.0 MeV and the range $a$ equals 1 fm.
Such parameters are suited to study the T-matrix in oscillator and momentum
representations both for bound and scattering states. Note that in the present
calculations the value of the depth does not play any significant role,
however it is important what sign it has or whether our potential is
attractive or repulsive.

\subsubsection{Extrapolation}
\label{Sec:Extrapol}

As pointed out above, diagonalization of the $N\times N$ matrix reveals
eigenfunctions $U_{n}^{\alpha }$ in a restricted range of the oscillator
quanta $0\leq n\leq N-1$. However, with a knowledge of the explicit form of the
expansions coefficients (\ref{eq:SP167}), one can extrapolate them to an
arbitrary value of $n$ beyond this area. To do this, we need to
determine the normalization factor $\mathcal{N}_{\alpha }$ and, what is  more
important, the discrete coordinate $r_{\alpha }$. This can be done, for
example, by solving a system of transcendent equations
\begin{eqnarray}
\mathcal{N}_{\alpha }\Phi _{N-2L}\left( r_{\alpha },b\right)
&=&U_{N-2}^{\alpha },  \label{eq:SP310} \\
\mathcal{N}_{\alpha }\Phi _{N-1L}\left( r_{\alpha },b\right)
&=&U_{N-1}^{\alpha },  \nonumber
\end{eqnarray}%
where the last two expansion coefficients $U_{N-2}^{\alpha }$ and $%
U_{N-1}^{\alpha }$ are determined by their predicted form (\ref{eq:SP167}). This set of
equations can be simplified. Assuming that $r_{\alpha }/b\ll R_N$,
we can refer to the asymptotic form (\ref{eq:SP170}) of the
oscillator functions
\begin{eqnarray}
\mathcal{N}_{\alpha }\left( -1\right) ^{N-2}\frac{2}{\sqrt{\pi }}\frac{1}{%
b^{3/2}}\sqrt{R_{N-2}}j_{L}\left( \frac{r_{\alpha }}{b}R_{N-2}\right)
&=&U_{N-2}^{\alpha },  \label{eq:SP312A} \\
\mathcal{N}_{\alpha }\left( -1\right) ^{N-1}\frac{2}{\sqrt{\pi }}\frac{1}{%
b^{3/2}}\sqrt{R_{N-1}}j_{L}\left( \frac{r_{\alpha }}{b}R_{N-1}\right)
&=&U_{N-1}^{\alpha }.  \label{eq:SP112B}
\end{eqnarray}%
Having obtained $\mathcal{N}_{\alpha }$ and $r_{\alpha }$ from
the suggested set of equations, we determine the eigenfunctions $U_{n}^{\alpha }$
 in an infinite range of $n$. We will denote the discrete
points $r_{\alpha }$ obtained from the numerical solution of Eqs.  (\ref%
{eq:SP310}) \ as $r_{\alpha }^{N}$. To verify that we correctly extrapolated
the eigenfunctions $U_{n}^{\alpha }$ to large values of $n$, we also
interpolate them to the region of small values of $n$ with $\mathcal{N}_{\alpha
}$ and $r_{\alpha }^{N}$ determined from the equations (\ref{eq:SP310}), and
then compare them with the exact one obtained by diagonalization of the
potential energy matrix. Approximate expansion coefficients valid in the
whole range of $n$ will be called the extrapolated eigenfunctions (coefficients). 
Besides,
with the determined value of $r_{\alpha }^{N}$ we can easily found the
position of a nearest node $N_{0}$ in the extrapolated region.

It is worthwhile noticing that the extrapolated expansion coefficients allow
one to determine eigenfunctions $\phi _{\alpha }\left( r,b\right) $ and $%
\phi _{\alpha }\left( p,b\right) $ in the whole region of the coordinate $r$
and momentum $p$, respectively. That might be important for investigating
the wave function (\ref{eq:SP303}) or T-matrix (\ref{eq:SP304}) at large
values of $r$ or $p$, respectively.

\subsubsection{Numerical illustrations}

\begin{figure}[ptbh]
\begin{center}
\includegraphics[
width=\textwidth]
{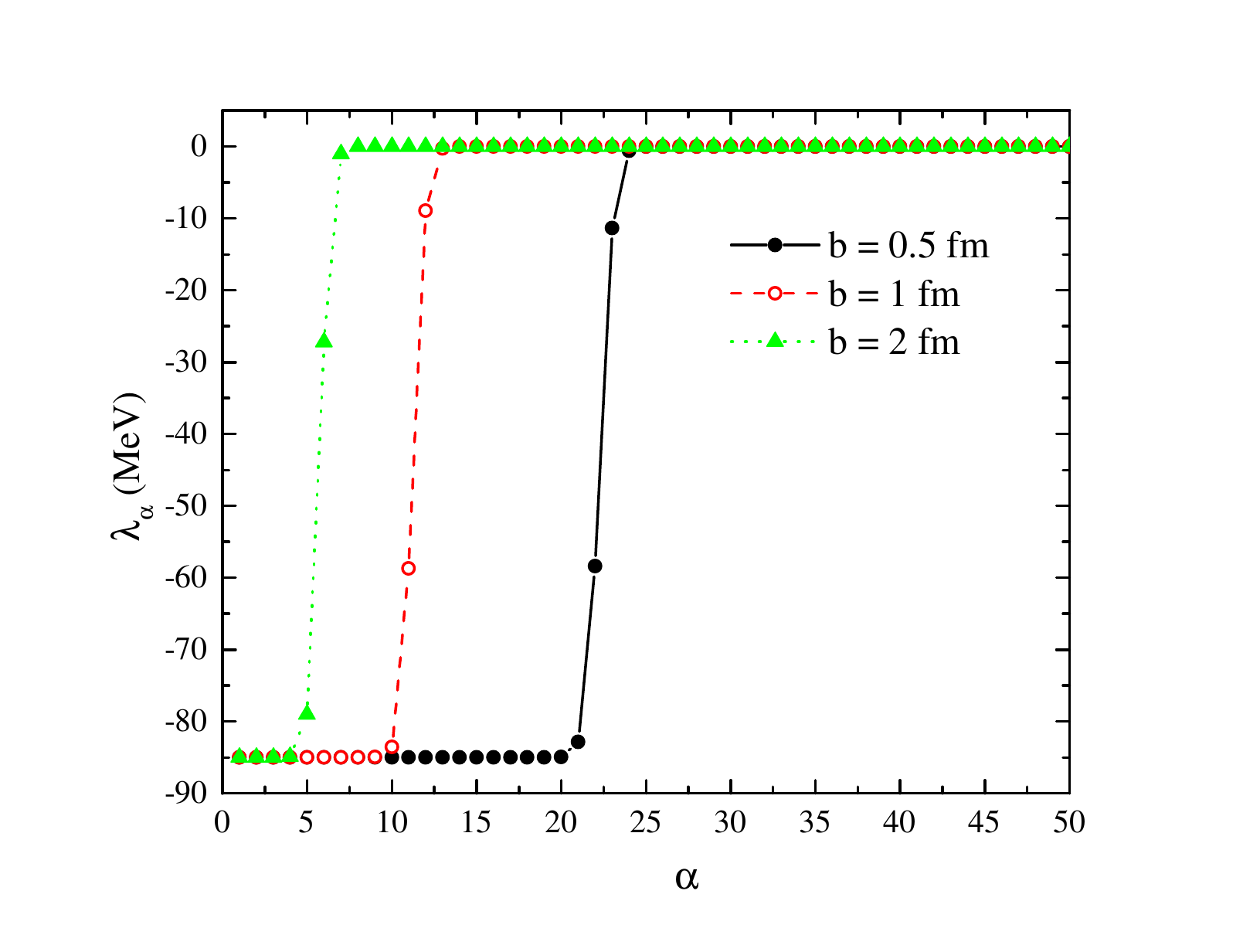}
\end{center}
\caption{Eigenvalues $\lambda_{\alpha}$ of the square well potential energy
operator versus the eigenvalue number $\alpha$ for three values of oscillator
length $b=0.5$ fm, $b=1$ fm and $b=2$ fm.}%
\label{Fig:V(alpha)_square}%
\end{figure}

\begin{figure}[ptbh]
\begin{center}
\includegraphics[
width=\textwidth]
{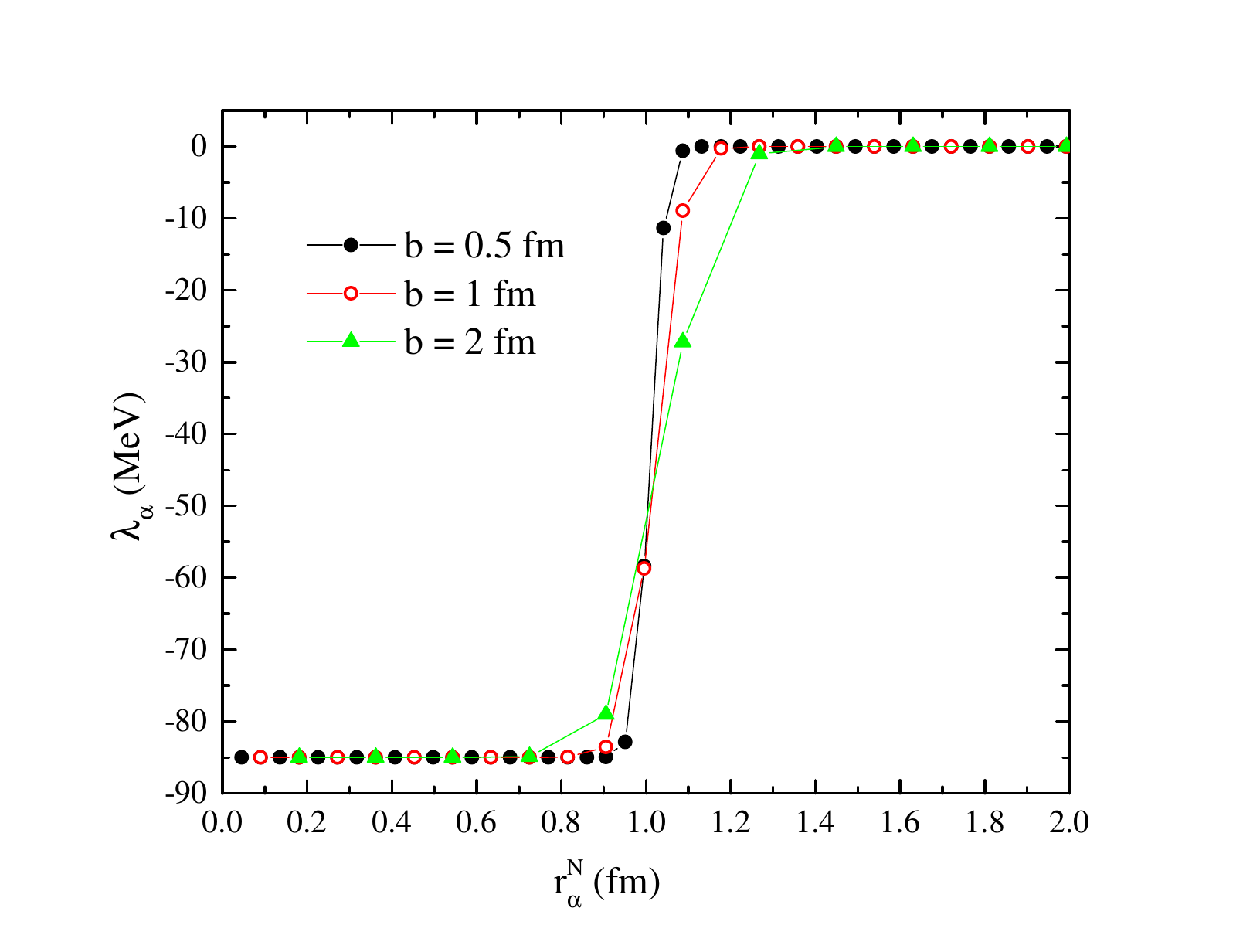}
\end{center}
\caption{Eigenvalues $\lambda_{\alpha}$ of the square well potential energy
operator versus a discrete coordinate $r_{\alpha}^{N}$ for three values of
oscillator length $b=0.5$ fm, $b=1$ fm and $b=2$ fm.}%
\label{Fig:V(r_a)_square}%
\end{figure}

\begin{figure}[ptbh]
\begin{center}
\includegraphics[
width=\textwidth]
{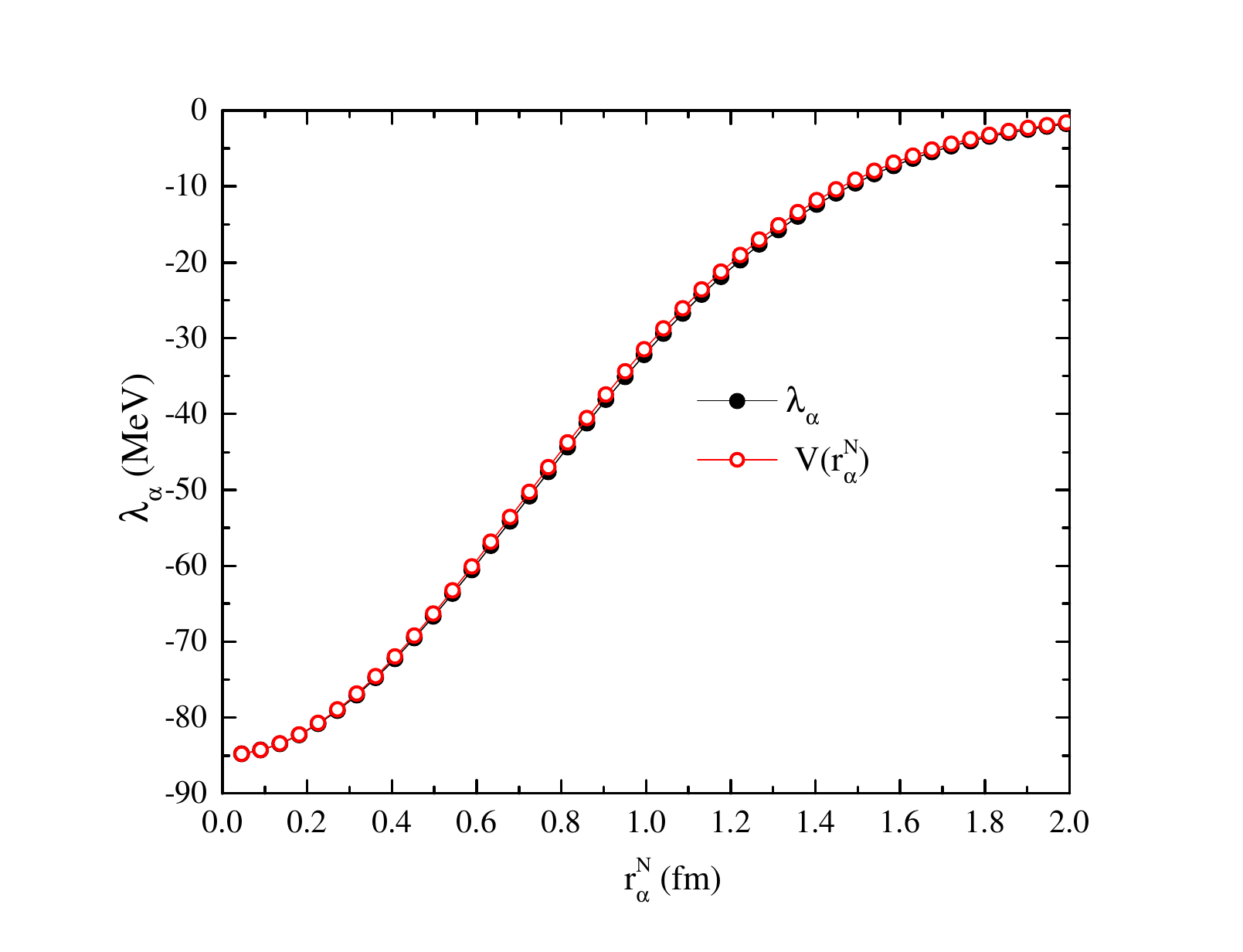}
\end{center}
\caption{Eigenvalues $\lambda_{\alpha}$ of the Gaussian potential energy
operator versus a discrete coordinate $r_{\alpha}^{N}$ for oscillator length
$b=0.5$ fm. Potential energy $V(r_{\alpha}^{N})$ in discrete points is given
by solid circles.}%
\label{Fig:V(r_a)_r0=05_gauss}%
\end{figure}

\begin{figure}[ptbh]
\begin{center}
\includegraphics[
width=\textwidth]
{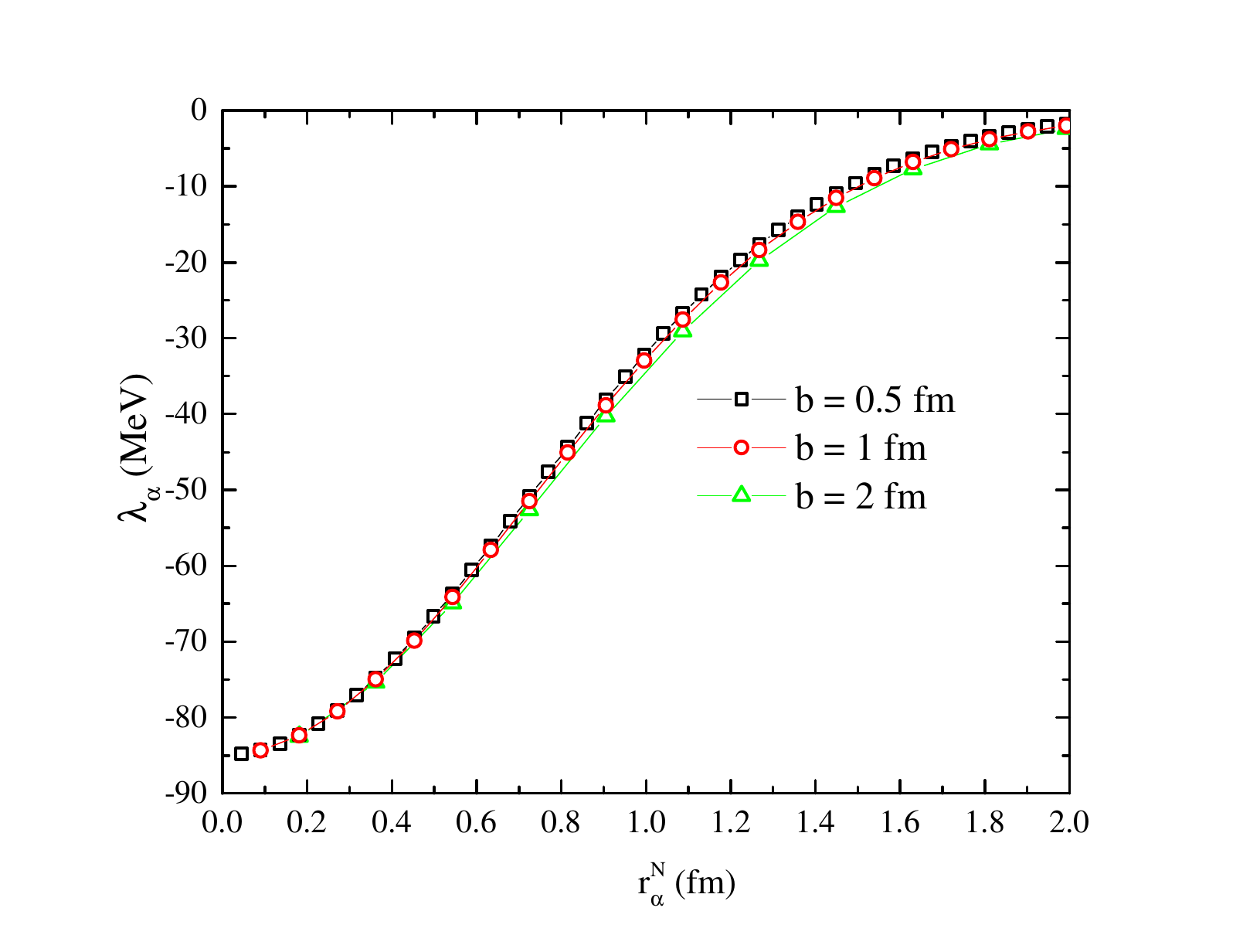}
\end{center}
\caption{Eigenvalues $\lambda_{\alpha}$ of the Gaussian potential energy
operator versus a discrete coordinate $r_{\alpha}^{N}$ for three values of
oscillator length $b=0.5$ fm, $b=1$ fm and $b=2$ fm.}%
\label{Fig:V(r_a)_gauss}%
\end{figure}

\begin{figure}[ptbh]
\begin{center}
\includegraphics[
width=\textwidth]
{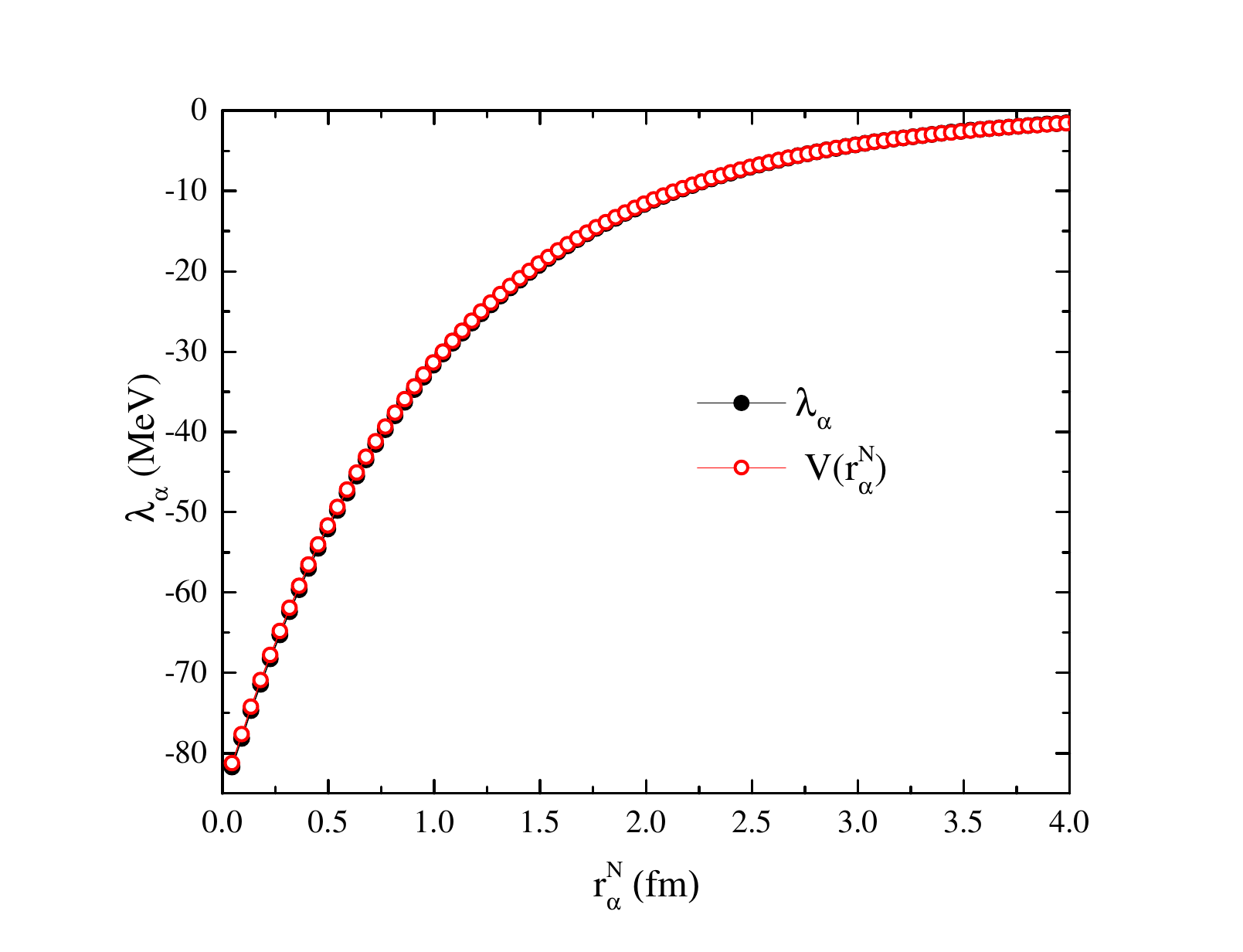}
\end{center}
\caption{Eigenvalues $\lambda_{\alpha}$ of the exponential potential energy
operator versus a discrete coordinate $r_{\alpha}^{N}$ for oscillator length
$b=0.5$ fm. Potential energy $V(r_{\alpha}^{N})$ in discrete points is given
by solid circles.}%
\label{Fig:V(r_a)_r0=05_exp}%
\end{figure}

\begin{figure}[ptbh]
\begin{center}
\includegraphics[
width=\textwidth]
{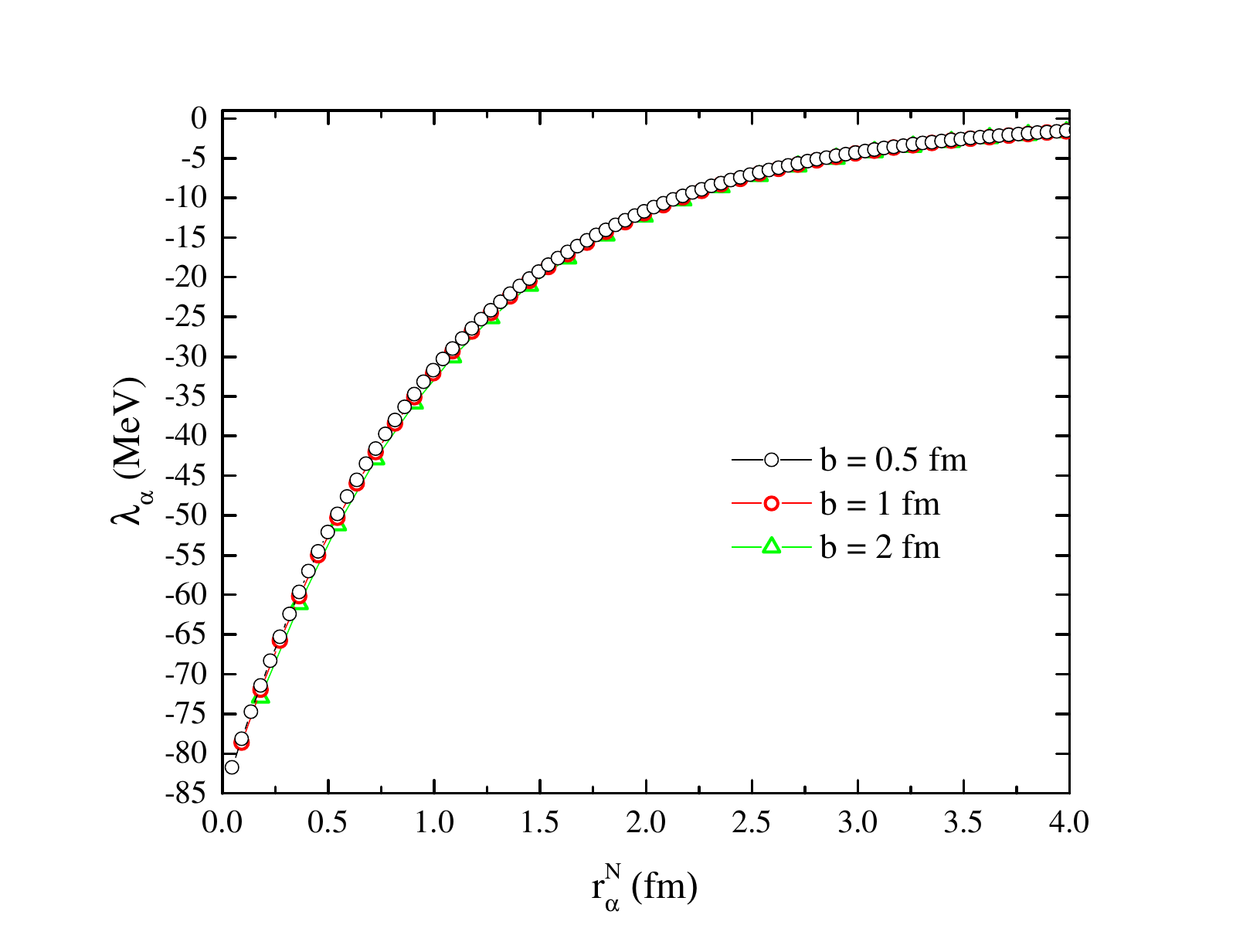}
\end{center}
\caption{Eigenvalues $\lambda_{\alpha}$ of the exponential potential energy
operator versus a discrete coordinate $r_{\alpha}^{N}$ for three values of
oscillator length $b=0.5$ fm, $b=1$ fm and $b=2$ fm.}%
\label{Fig:V(r_a)_exp}%
\end{figure}

\begin{figure}[ptbh]
\begin{center}
\includegraphics[
width=\textwidth]
{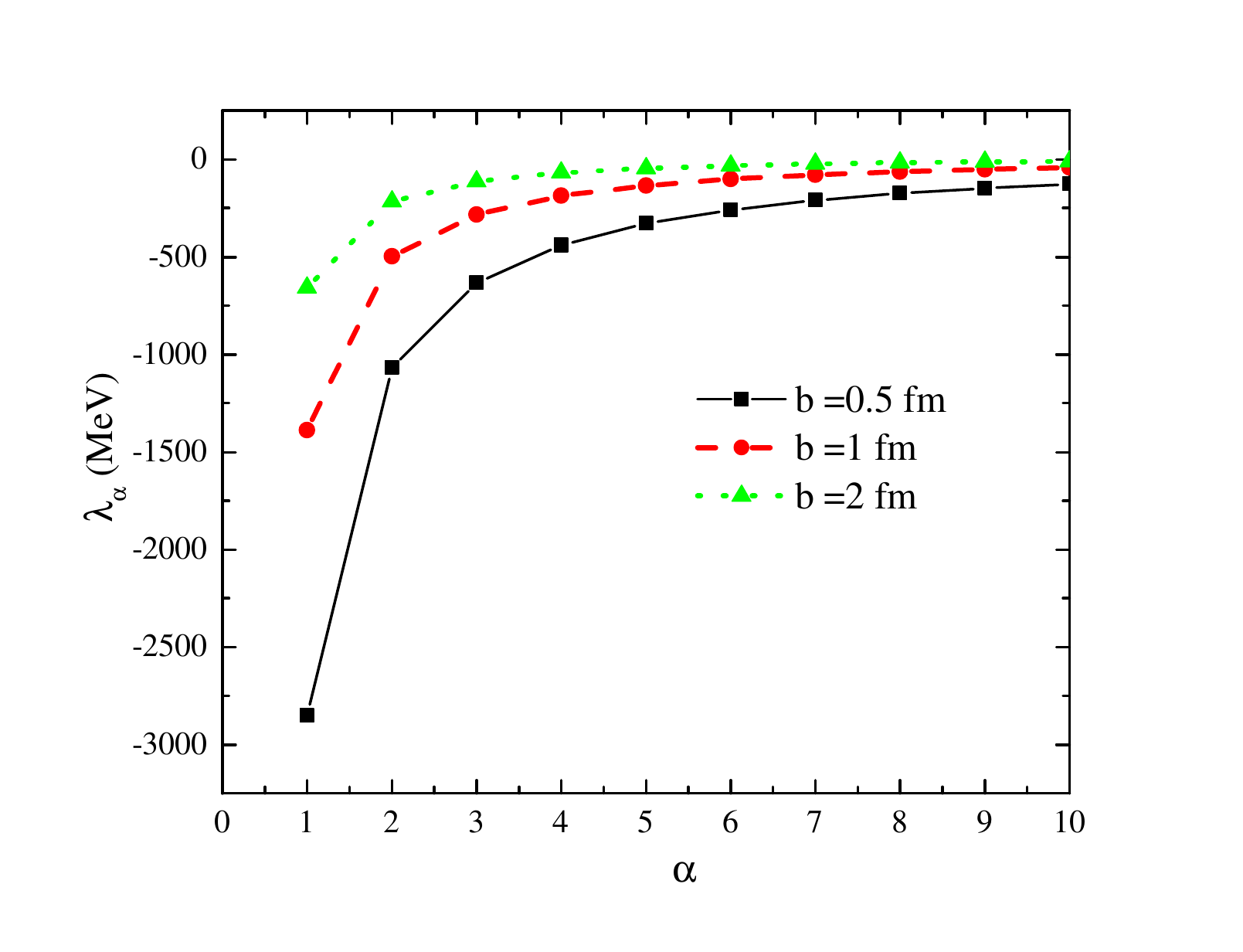}
\end{center}
\caption{Eigenvalues $\lambda_{\alpha}$ of the Yukawa potential energy
operator versus a number of eigenvalue $\alpha$ for three values of oscillator
length $b=0.5$ fm, $b=1$ fm and $b=2$ fm.}%
\label{Fig:V(alpha)_yukawa}%
\end{figure}

\begin{figure}[ptbh]
\begin{center}
\includegraphics[
width=\textwidth]
{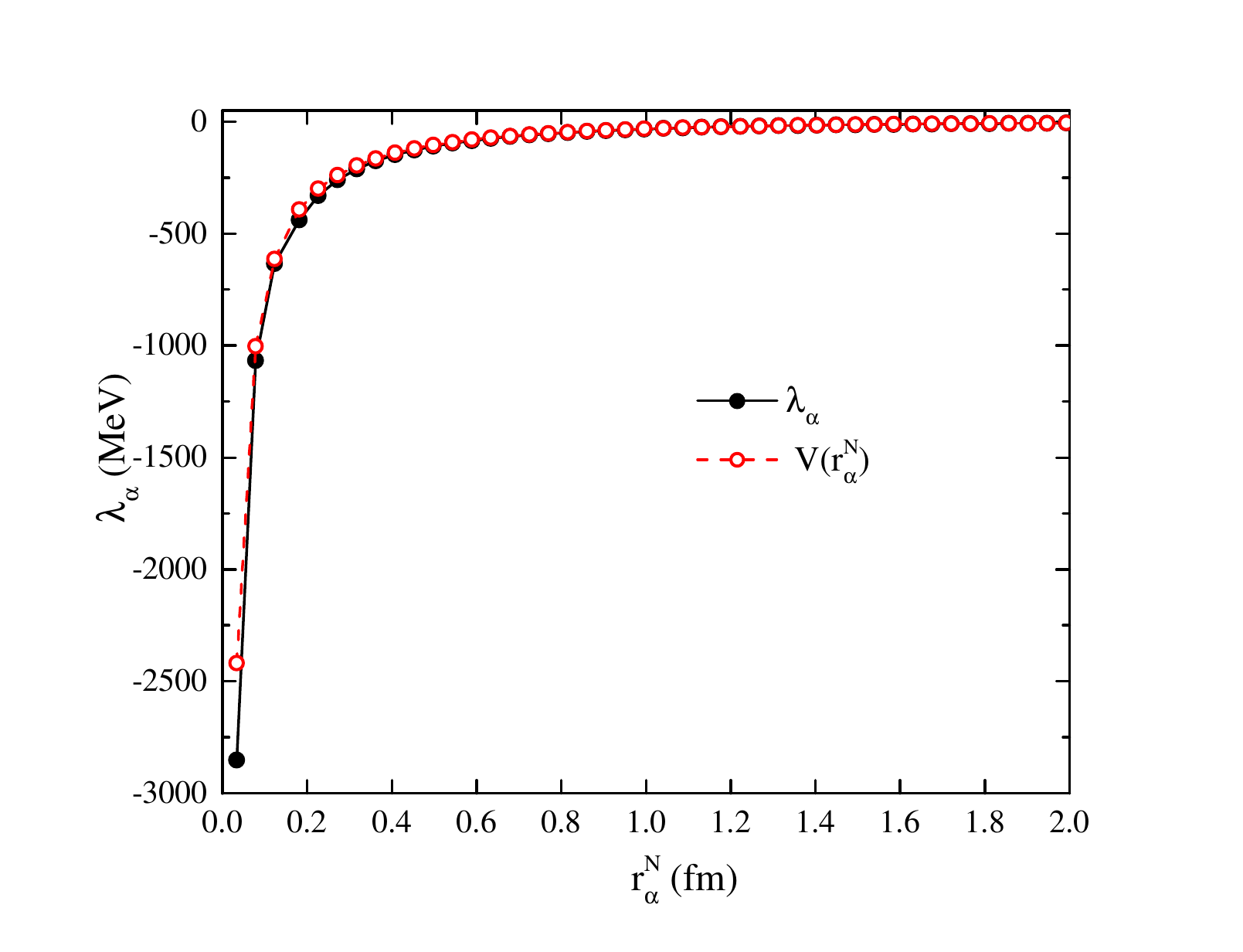}
\end{center}
\caption{Eigenvalues $\lambda_{\alpha}$ of the Yukawa potential energy
operator versus a discrete coordinate $r_{\alpha}^{N}$ for oscillator length
$b=0.5$ fm. Potential energy $V(r_{\alpha}^{N})$ in discrete points is given
by solid circles.}%
\label{Fig:V(r_a)_r0=05_yukawa}%
\end{figure}

\begin{figure}[ptbh]
\begin{center}
\includegraphics[
width=\textwidth]
{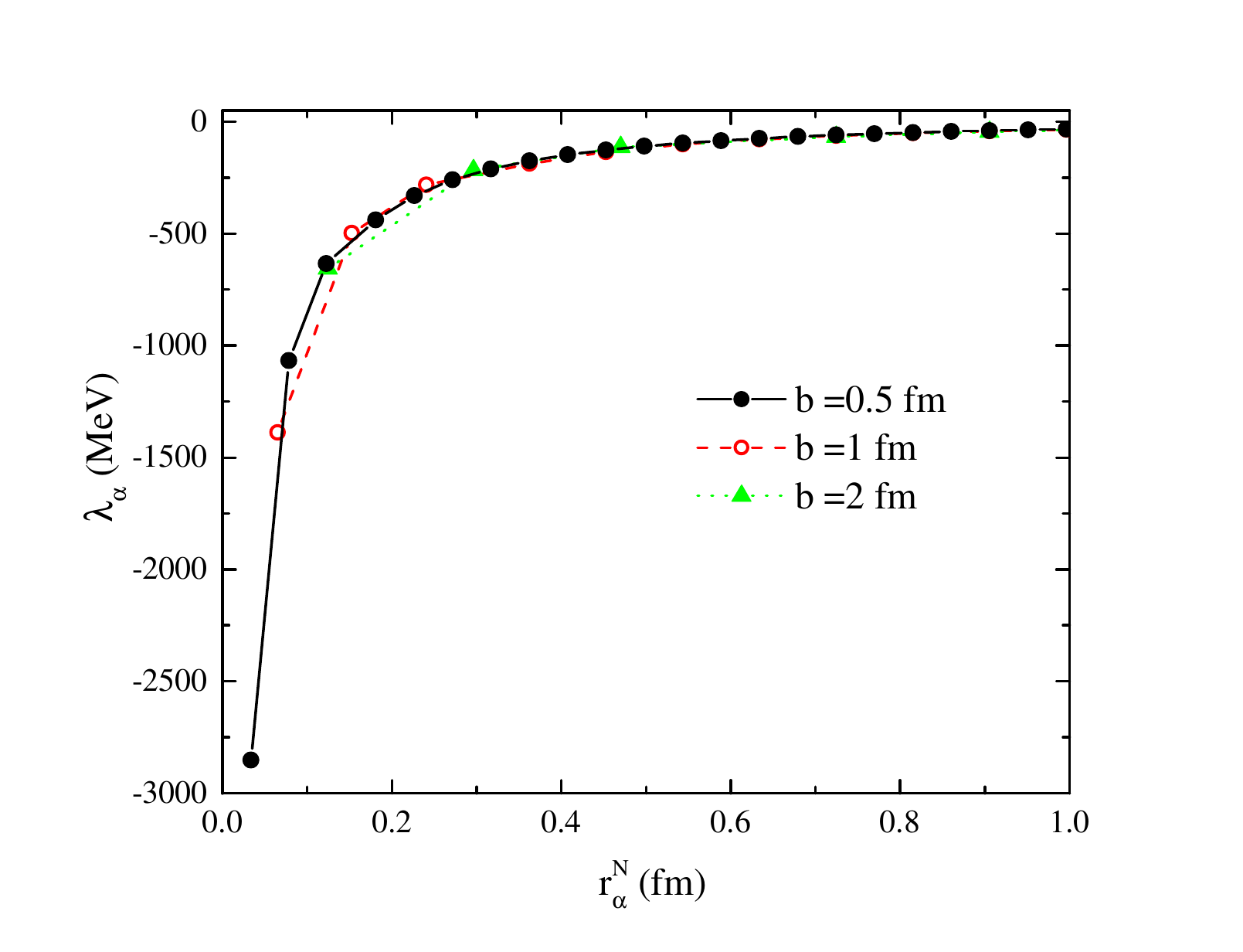}
\end{center}
\caption{Eigenvalues $\lambda_{\alpha}$ of the Yukawa potential energy
operator versus a discrete coordinate $r_{\alpha}^{N}$ for three values of
oscillator length $b=0.5$ fm, $b=1$ fm and $b=2$ fm.}%
\label{Fig:V(r_a)_yukawa}%
\end{figure}

\begin{figure}[ptbh]
\begin{center}
\includegraphics[
width=\textwidth]
{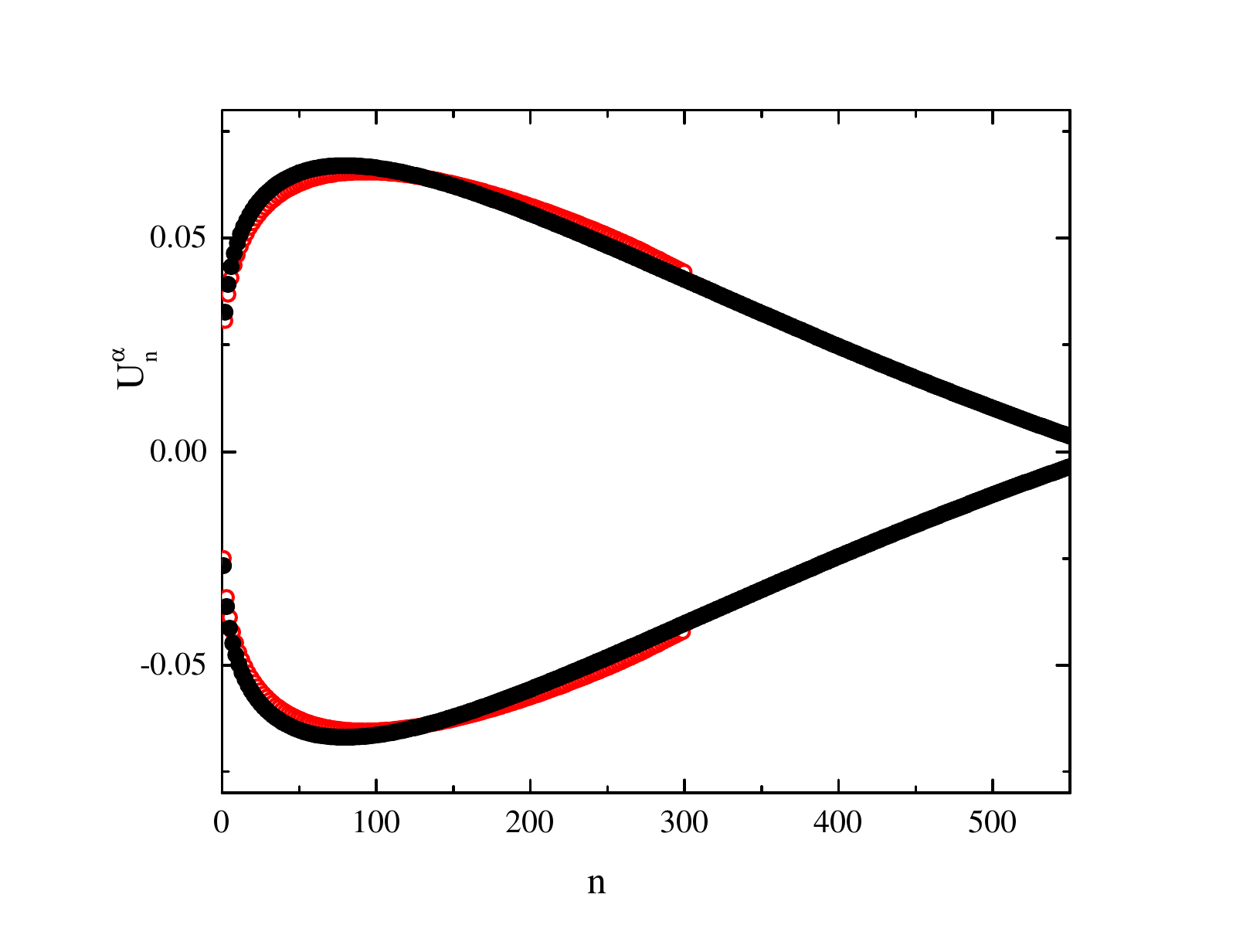}
\end{center}
\caption{Eigenfunction $U^{\alpha}_{n},$ $\alpha=1$ of the Yukawa potential
energy operator in the discrete representation versus the number of oscillator
quanta $n$ for oscillator length $b=1$ fm. Extrapolated eigenfunction 
(Oscillator function $\mathcal{N}_{\alpha }\Phi_{n}%
(r^{N}_{\alpha},b)$) is given by solid circles.}%
\label{Fig:V(n)_alpha=1_r0=1_yukawa}%
\end{figure}

\begin{figure}[ptbh]
\begin{center}
\includegraphics[
width=\textwidth]
{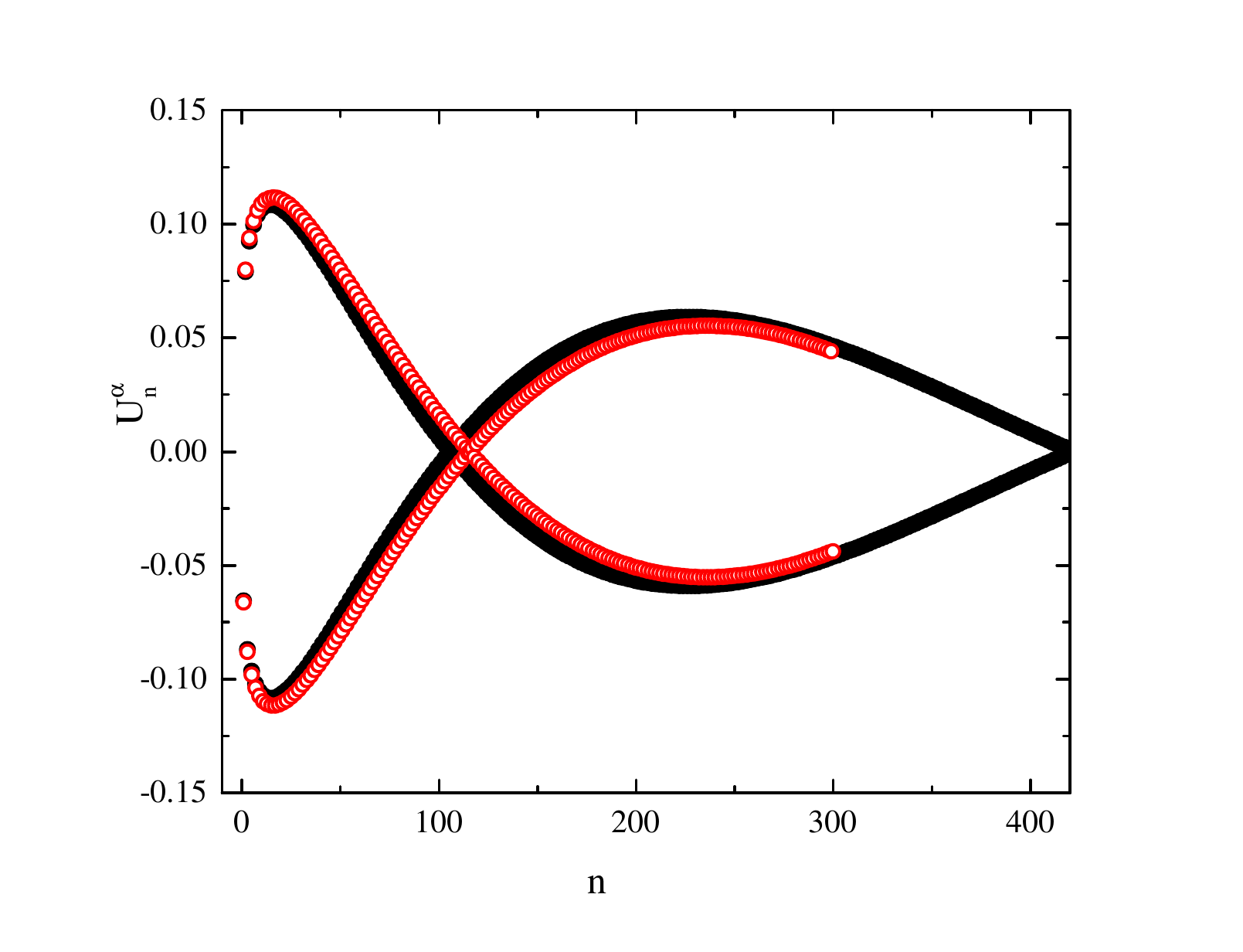}
\end{center}
\caption{Eigenfunction $U^{\alpha}_{n},$ $\alpha=2$ of the Yukawa potential
energy operator in the discrete representation versus the number of oscillator
quanta $n$ for oscillator length $b=1$ fm. Extrapolated eigenfunction 
(Oscillator function $\mathcal{N}_{\alpha }\Phi_{n}%
(r^{N}_{\alpha},b)$) is given by solid circles.}%
\label{Fig:V(n)_alpha=2_r0=1_yukawa}%
\end{figure}

\begin{figure}[ptbh]
\begin{center}
\includegraphics[
width=\textwidth]
{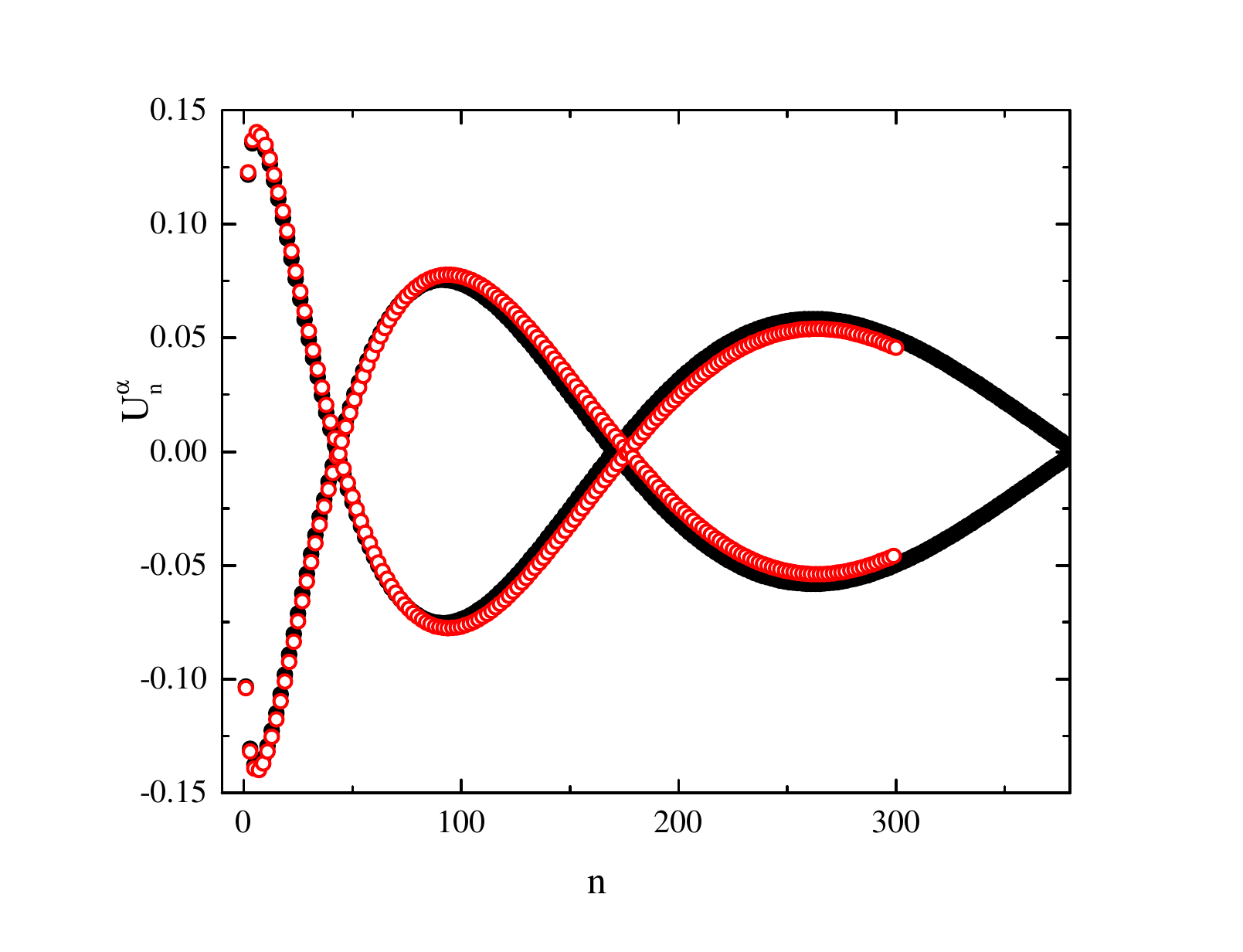}
\end{center}
\caption{Eigenfunction $U^{\alpha}_{n},$ $\alpha=3$ of the Yukawa potential
energy operator in the discrete representation versus the number of oscillator
quanta $n$ for oscillator length $b=1$ fm. Extrapolated eigenfunction 
(Oscillator function $\mathcal{N}_{\alpha }\Phi_{n}%
(r^{N}_{\alpha},b)$) is given by solid circles.}%
\label{Fig:V(n)_alpha=3_r0=1_yukawa}%
\end{figure}

\begin{figure}[ptbh]
\begin{center}
\includegraphics[
width=\textwidth]
{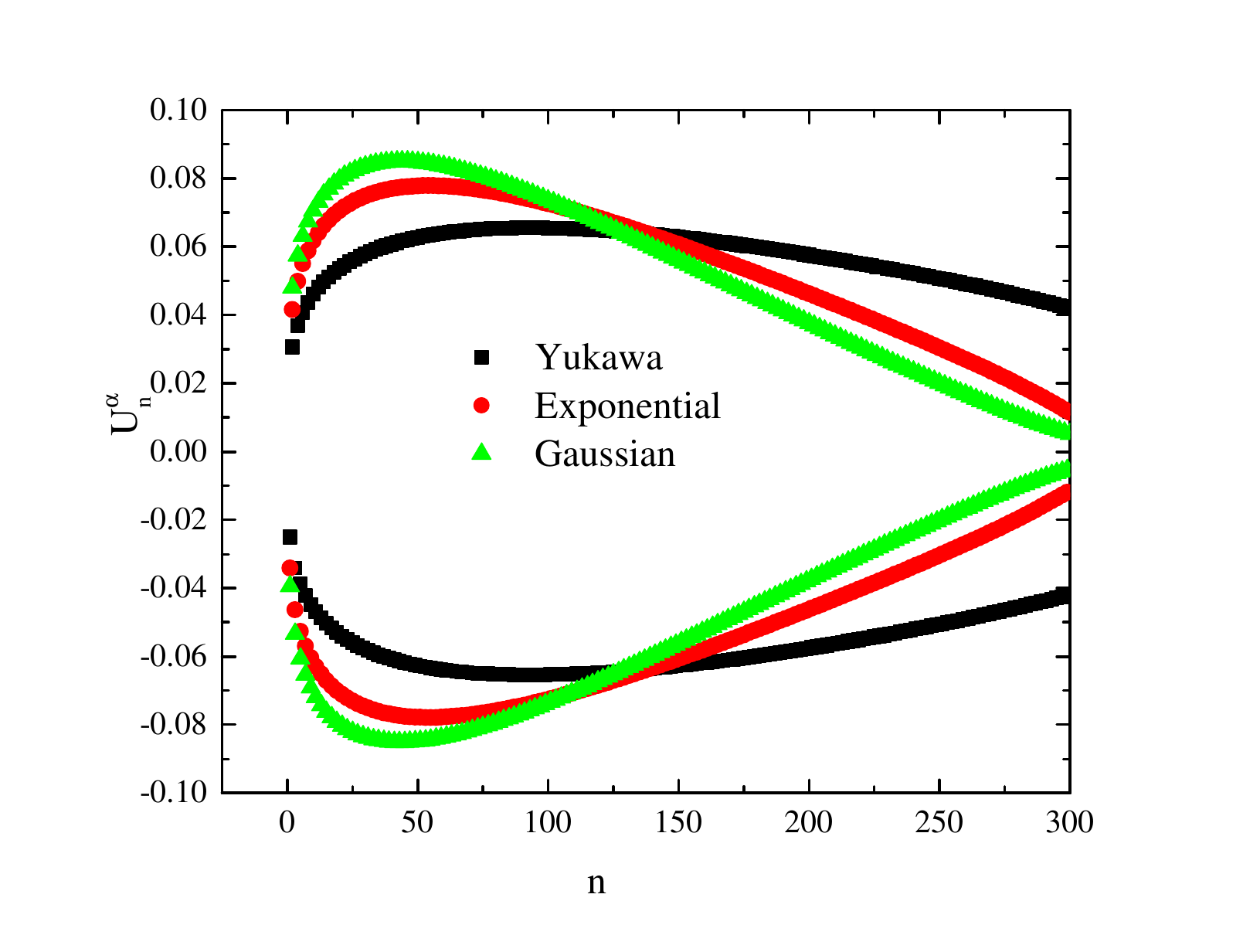}
\end{center}
\caption{Eigenfunctions $U^{\alpha}_{n},$ $\alpha=1$ of the potential energy
operator for different potentials in the discrete representation versus the
number of oscillator quanta $n$ for oscillator length $b=1$ fm. The types of
the potentials are shown in the figure.}%
\label{Fig:V(n)_alpha=1_r0=1}%
\end{figure}

\begin{figure}[ptbh]
\begin{center}
\includegraphics[
width=\textwidth]
{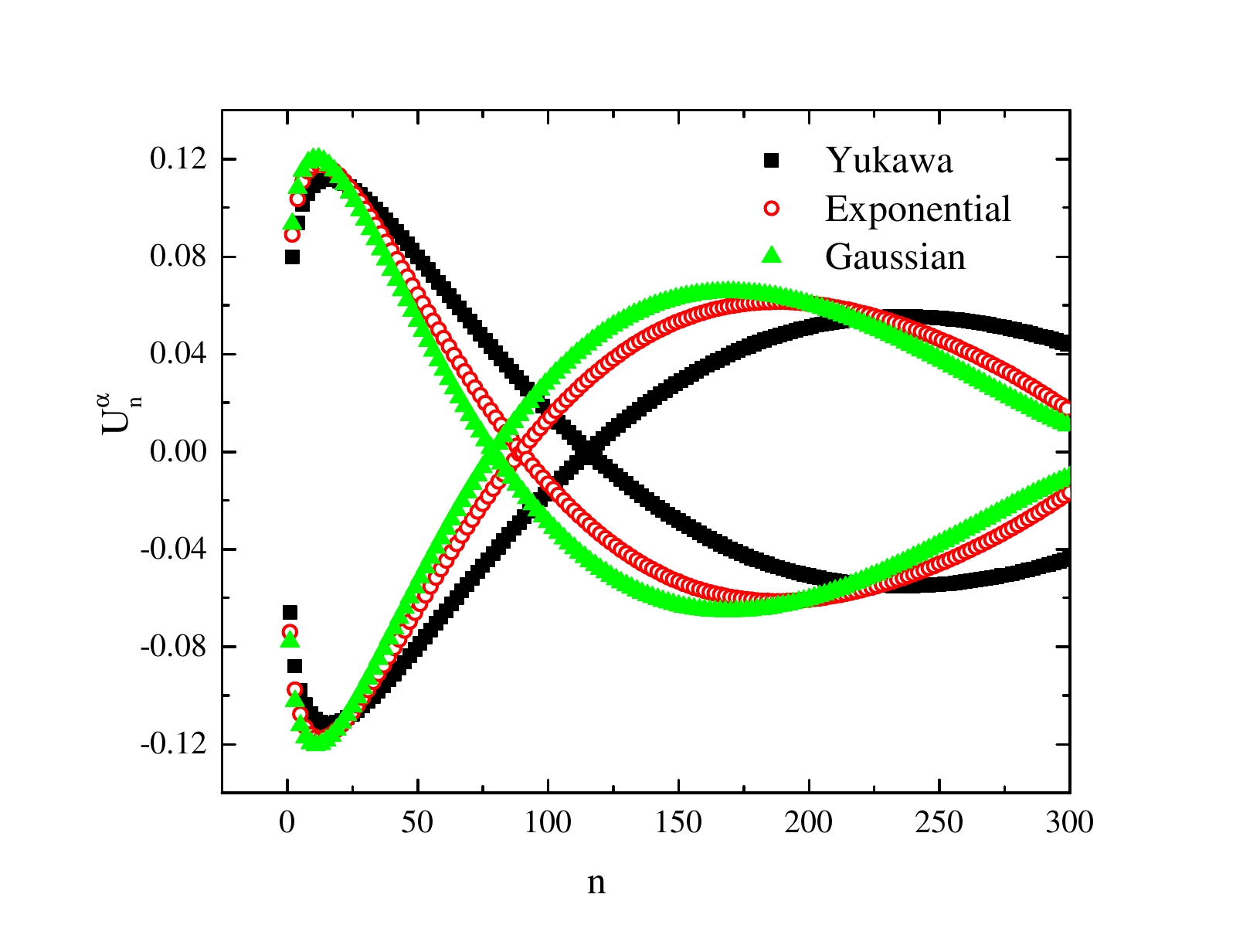}
\end{center}
\caption{Eigenfunctions $U^{\alpha}_{n},$ $\alpha=2$ of the potential energy
operator for different potentials in the discrete representation versus the
number of oscillator quanta $n$ for oscillator length $b=1$ fm. The types of
the potentials are shown in the figure.}%
\label{Fig:V(n)_alpha=2_r0=1}%
\end{figure}

In Figures (\ref{Fig:V(alpha)_square})--(\ref{Fig:V(p)_alpha_r0=1_yukawa})
we consider the main
properties of eigenvalues and eigenfunctions of the potential energy matrix in
oscillator space.

Figure \ref{Fig:V(alpha)_square} illustrates the behavior of the eigenvalues
$\lambda_{\alpha}$ of the square well potential energy operator against the
eigenvalue number $\alpha$. This dependence is demonstrated for three values of oscillator length $b$.
The first thing to meet the eye is reproducing the shape of the potential by
the eigenvalues $\lambda_{\alpha}$. The depth of the square well potential in
the representation of its eigenvalues is just the same as the intensity of the
original potential energy operator, while the range of the potential is seemed
to depend on the value of oscillator length. Indeed, as can be seen from Fig.
\ref{Fig:V(alpha)_square}, the larger is the oscillator length, the smaller is
the width of the potential well.

However, it is a superficial difference. If we plot the eigenvalues
$\lambda_{\alpha}$ against a discrete coordinate $r_{\alpha}^{N}$ (see Fig.
\ref{Fig:V(r_a)_square}), we will find that the depth and the range of the
square well potential in the representation of its eigenvalues are the same
for all the values of the oscillator length and coincide with those of the
original potential. The only difference is that the smaller is the oscillator
length, the smaller is the increment of a discrete variable $r_{\alpha}^{N}$.

Important information can be deduced from Fig. \ref{Fig:V(alpha)_square}.
We see that the large number of eigenvalues equal  zero. The number of nonzero eigenvalues
strongly depends on the oscillator length $b$. The smaller is $b$, the larger
is the number of the eigenvalues $\lambda _{\alpha }\neq 0$.  For all three values of $b$,
the number of the non-vanishing eigenvalues  is not more than 10\%
of the total number of the obtained eigenvalues. Thus, we can conclude that the
real number of eigenvalues involved in Eqs. (\ref{eq:SP303}) and (\ref%
{eq:SP304}) to form wave function and T-matrix is very small.  Similar
situation is observed for other potentials.  Due to relatively long
tails of these potentials compared with the square-well potential, the
number of the nonzero eigenvalues is larger, however the number of
eigenvalues $\lambda _{\alpha }=0$ is even greater.

Fig. \ref{Fig:V(r_a)_r0=05_gauss} shows perfect coincidence of the potential
energy $V(r_{\alpha}^{N})$ in discrete points $r_{\alpha}^{N}$ with
eigenvalues $\lambda_{\alpha}$ for the oscillator length $b=0.5$ fm, which is
a half the range of the potential.

Fig. \ref{Fig:V(r_a)_gauss} demonstrates that small values of oscillator
length scan the potential in more detail. 

Fig. \ref{Fig:V(r_a)_r0=05_exp} and Fig. \ref{Fig:V(r_a)_exp} evidence that
our conclusions are valid equally well for a Gaussian potential.

Figure \ref{Fig:V(alpha)_yukawa} represents the dependence of the eigenvalues
of the Yukawa potential energy operator on the eigenvalue number and
oscillator length. We have chosen this potential to check how the shape of a
singular at zero potential is reproduced by its eigenvalues $\lambda_{\alpha}%
$. As can be observed from Fig. \ref{Fig:V(alpha)_yukawa}, quite small values
of oscillator length are needed to reach an interior part of the potential.

Fig. \ref{Fig:V(r_a)_r0=05_yukawa} demonstrates that Yukawa potential is also
well reproduced by the behavior of its eigenvalues, although the coincidence
with the original potential at very small values of a discrete variable is not
so ideal as for a non-singular potential.

At the same time, Fig. \ref{Fig:V(r_a)_yukawa} demonstrates the same
regularity as the other potentials, namely, the more detailed scanning of the
shape of the potential by smaller values of oscillator length.

Eigenfunctions $U^{\alpha}_{n}$ of the Yukawa potential energy operator in the
harmonic oscillator representation corresponding to the three lowest
eigenvalues $\lambda_{\alpha},$ $\alpha=1,2,3$ are plotted against the number
of oscillator quanta $n$ in Figs.\ref{Fig:V(n)_alpha=1_r0=1_yukawa},
\ref{Fig:V(n)_alpha=2_r0=1_yukawa}, \ref{Fig:V(n)_alpha=3_r0=1_yukawa}.
Oscillator length $b$ was chosen to be equal to the range of the potential.

In Figs.\ref{Fig:V(n)_alpha=1_r0=1_yukawa}, \ref{Fig:V(n)_alpha=2_r0=1_yukawa}%
, \ref{Fig:V(n)_alpha=3_r0=1_yukawa} we can observe that eigenfunctions
$U^{\alpha}_{n}$ are oscillator functions $\Phi_{n}(r^{N}_{\alpha},b)$, as was
concluded in the previous Section of the paper. Predictably, in the range from
zero to $N_{max}$ eigenfunction $U^{\alpha=1}_{n}$ is nodeless, while
$U^{\alpha=2}_{n}$ has a node, and $U^{\alpha=3}_{n}$ has two nodes.

Figures \ref{Fig:V(n)_alpha=1_r0=1} and \ref{Fig:V(n)_alpha=2_r0=1} present
eigenfunctions $U^{\alpha}_{n},$ $\alpha=1, 2$ for a Gaussian, an exponential,
and Yukawa potentials. In all the cases the oscillator length is equal to the
range of the potential. Figures \ref{Fig:V(n)_alpha=1_r0=1},
\ref{Fig:V(n)_alpha=2_r0=1} demonstrate close resemblance of the
eigenfunctions which proves their universality. Indeed, all the eigenfunctions
are the harmonic oscillator functions corresponding to the same value of the
oscillator length. The difference between the eigenfunctions for different
potentials is due to the distinction between the corresponding discrete
coordinates $r^{N}_{\alpha}.$ In turn, value of $r^{N}_{\alpha}$ is determined
by the location of the $\alpha$th node of the eigenfunction, i.e., the number
of oscillator quanta $N$.

As can be observed from Fig. \ref{Fig:V(n)_alpha=1_r0=1}, the eigenfunction of
the Yukawa potential has the first node much farther than the eigenfunctions
of two other potentials. That is why the distinction between the eigenfunctions of
a Gaussian and an exponential potential is less than the difference between
the latter eigenfunctions and the eigenfunction of the Yukawa potential.

\subsubsection{Comparison}

There is another way to demonstrate the validity of the expression (\ref%
{eq:SP167}) for the eigenfunctions of the potential energy matrix in the oscillator
representation. Suppose we obtained eigenfunctions and eigenvalues of the
same potential with two different numbers of oscillator functions $N_{1}$ and
$N_{2}$, with $N_{1}<N_{2}$. And suppose that the eigenvalue $\lambda
_{\alpha _{1}}$ obtained with  $N_{1}$ oscillator functions approximately
equals the eigenvalue $\lambda _{\alpha _{2}}$ calculated with $N_{2}$
oscillator functions. Then the eigenfunctions $U_{n}^{\alpha _{1}}$ and $%
U_{n}^{\alpha _{2}}$ will have the similar behavior in the common range of $%
n $: $0\leq n\leq N_{1}-1$. They have approximately the same value of $%
r_{\alpha }$, but differ in the normalization constants $\mathcal{N}_{\alpha
_{1}}$and $\mathcal{N}_{\alpha _{2}}$. If the eigenfunctions $U_{n}^{\alpha_{1}}$ have one
or more nodes in the range $0\leq n\leq N_{1}-1$, the eigenfunction
$U_{n}^{\alpha_{2}}$ should have node(s) at the same point(s). By using a
common normalization constant, for example $\mathcal{N}_{\alpha_{1}}$, for
both eigenfunctions $U_{n}^{\alpha_{1}}$ and $U_{n}^{\alpha_{2}}$, we will see
that they totally coincide in the range $0\leq n\leq N_{1}-1$.

\begin{figure}
[ptb]
\begin{center}
\includegraphics[
width=\textwidth]%
{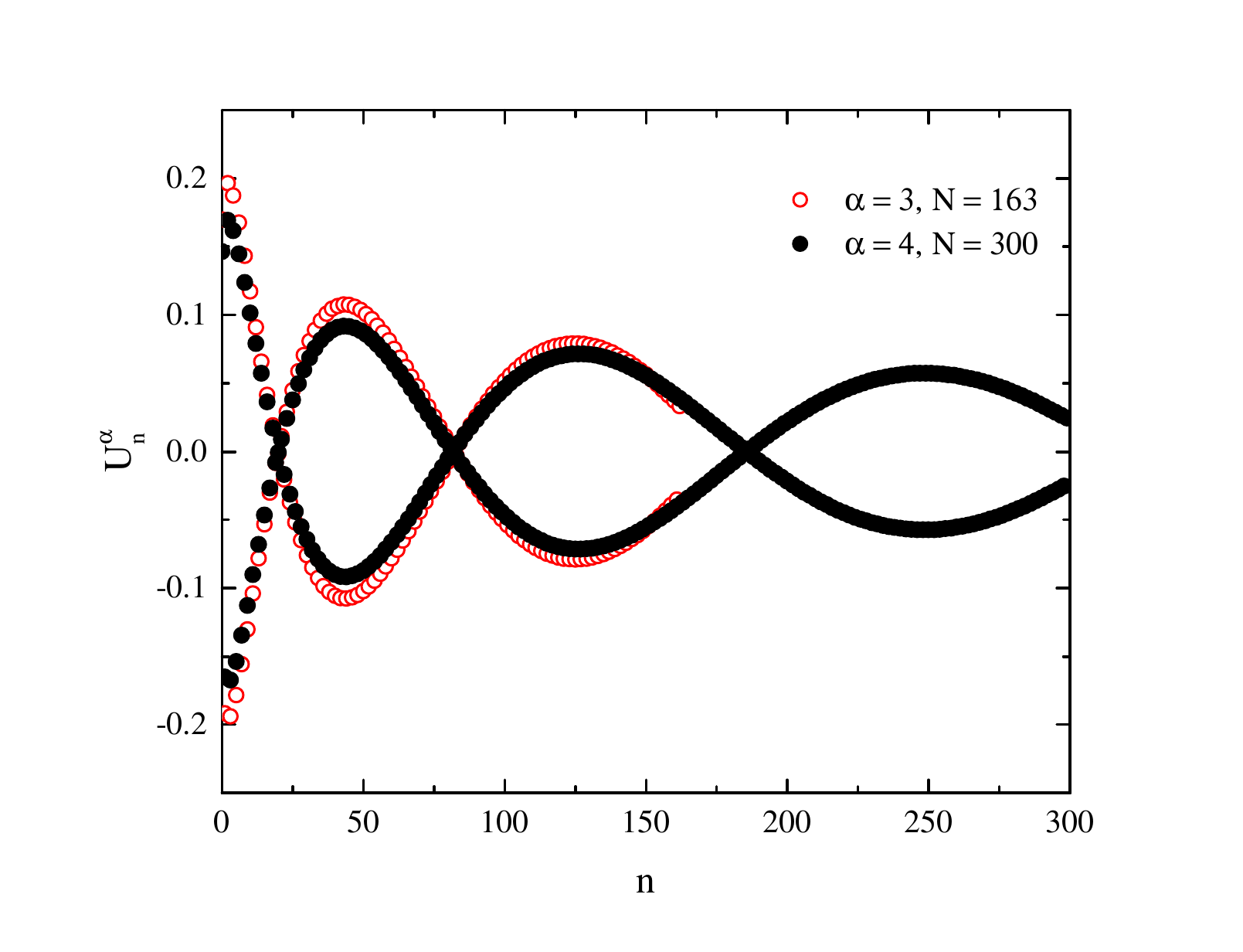}%
\caption{Eigenfunctions of the potential energy matrix in the oscillator
representation obtained for the
exponential potential with $b$=1 fm and two different numbers of the oscillator
functions.}%
\label{Fig:EignFunsExpComp}%
\end{center}
\end{figure}

This statement is demonstrated in Fig. \ref{Fig:EignFunsExpComp} for the
exponential potential. These results are obtained with the oscillator length
$b$= 1 fm. Two eigenfunctions displayed in Fig. \ref{Fig:EignFunsExpComp} are
obtained with $N_{1}$=163 and $N_{2}$=300 oscillator functions. We compare
eigenfunctions for the $\alpha_{1}=3$ and $\alpha_{2}=4$ eigenstates which
have very close eigenvalues $\lambda_{\alpha_{1}}\approx\lambda_{\alpha_{2}}$.
As we see, these two eigenfunctions have a similar behavior in the range $0\leq
n\leq162$, and besides they have two nodes at the same value of $n$. We do not
renormalize them in the same fashion, thus they slightly differ.%

\begin{figure}[ptbh]
\begin{center}
\includegraphics[
width=\textwidth]
{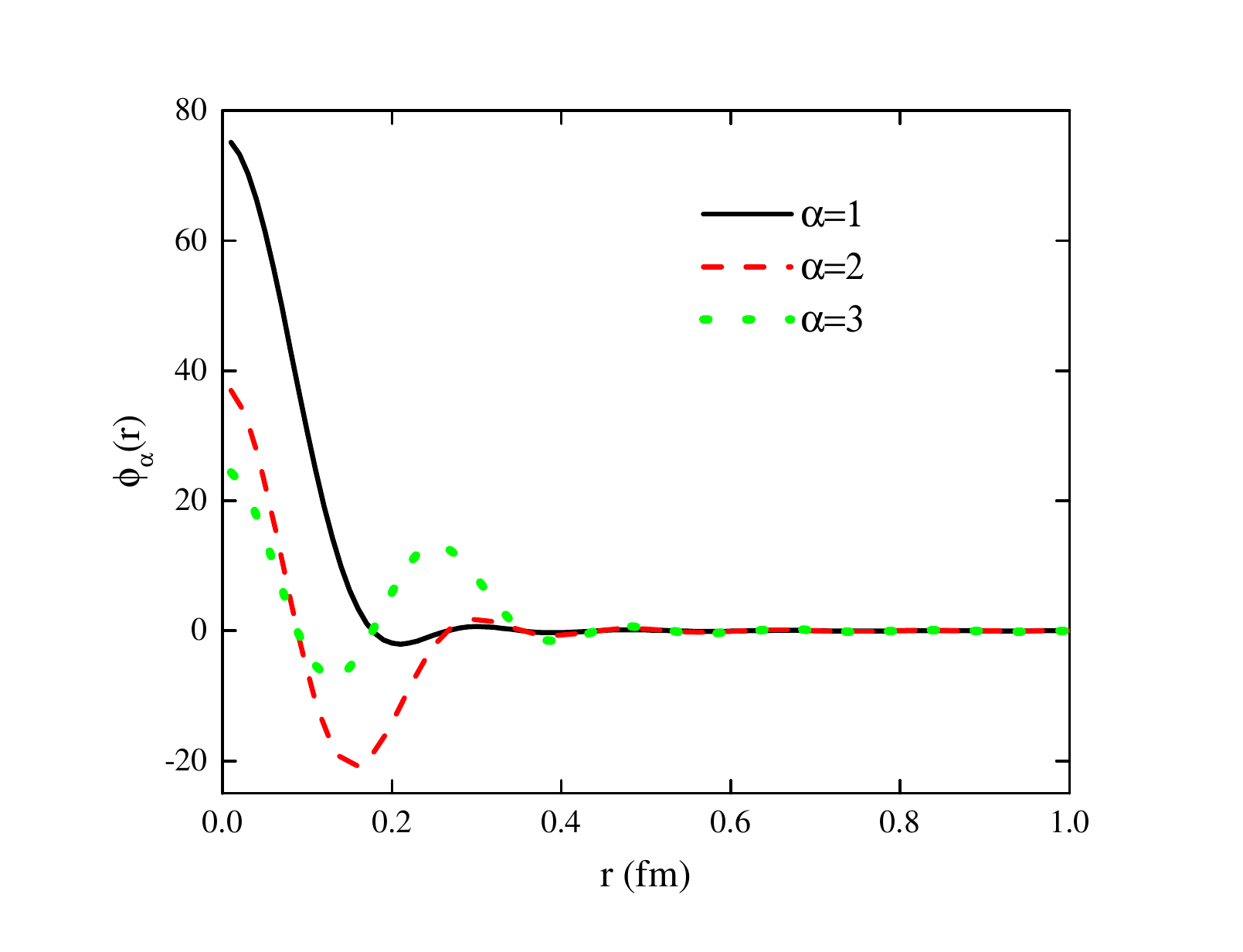}
\end{center}
\caption{Eigenfunctions $\phi_{\alpha}(r)$ of the Gaussian potential energy
operator in the coordinate representation for oscillator length $b=1$ fm.
Values of $\alpha$ are indicated in the figure.}%
\label{Fig:V(r)_alpha_r0=1_gauss}%
\end{figure}

\begin{figure}[ptbh]
\begin{center}
\includegraphics[
width=\textwidth]
{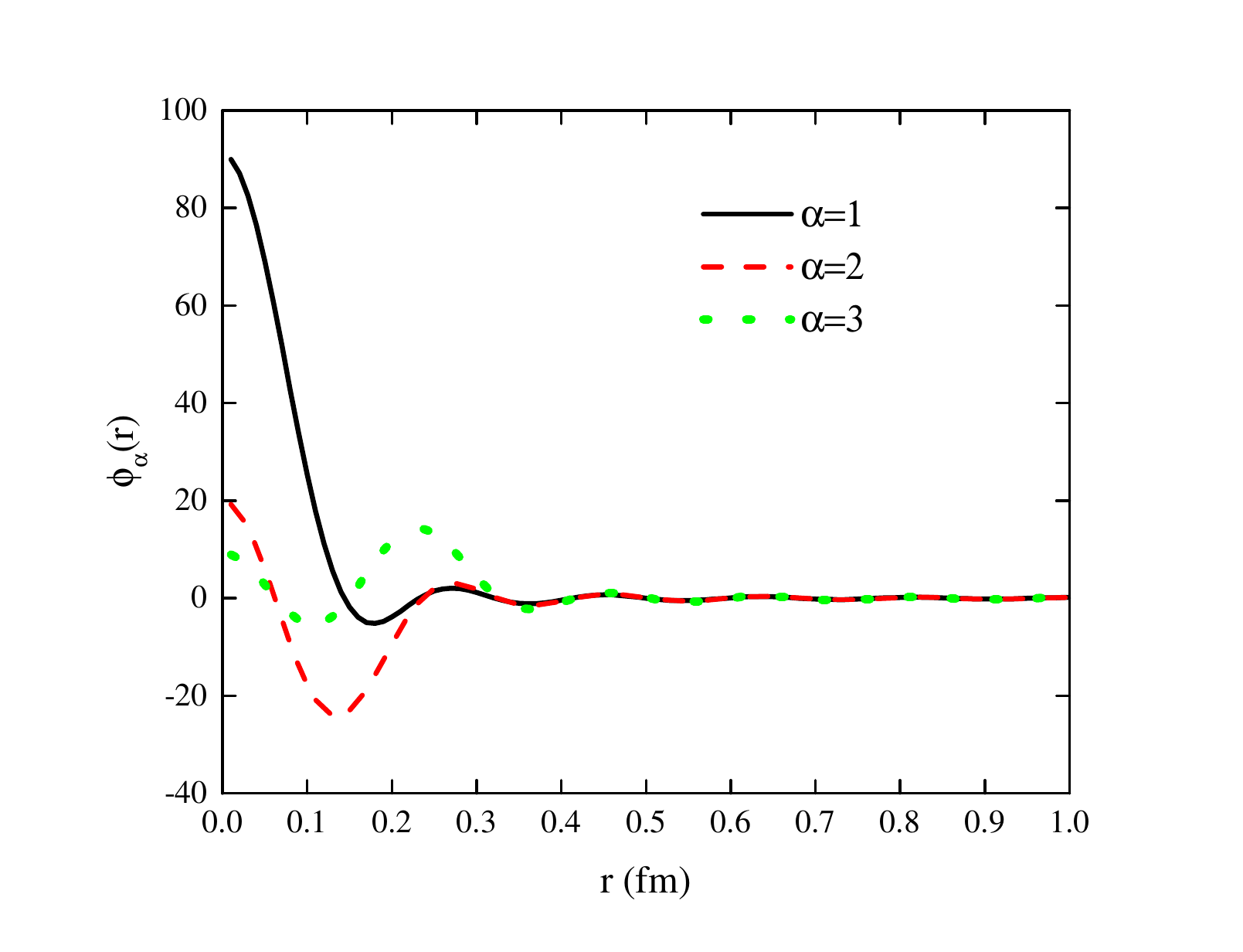}
\end{center}
\caption{Eigenfunctions $\phi_{\alpha}(r)$ of the Yukawa potential energy
operator in the coordinate representation for oscillator length $b=1$ fm.
Values of $\alpha$ are indicated in the figure.}%
\label{Fig:V(r)_alpha_r0=1_yukawa}%
\end{figure}

Fig. \ref{Fig:V(r)_alpha_r0=1_gauss} and Fig. \ref{Fig:V(r)_alpha_r0=1_yukawa}
depict first three eigenfunctions $\phi_{\alpha}(r)$ of the Gaussian and Yukawa potentials in
coordinate representation. We can observe that the smaller is the eigenvalue
number $\alpha$, the more the corresponding eigenfunction resembles the delta
function. It might be well to point out that the number of nodes in the
eigenfunctions increases with increasing the eigenvalue number $\alpha.$

\begin{figure}[ptbh]
\begin{center}
\includegraphics[
width=\textwidth]
{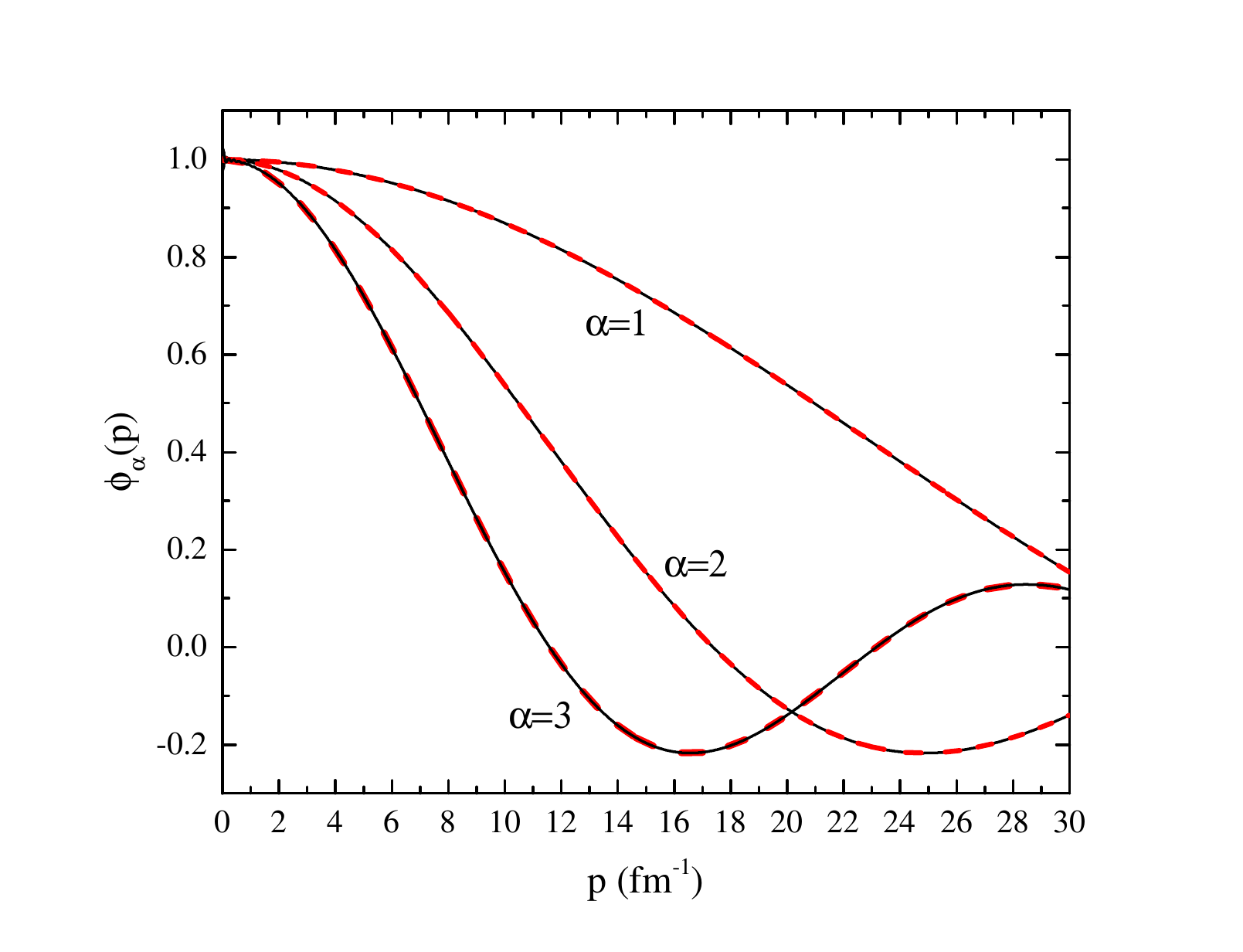}
\end{center}
\caption{Eigenfunctions $\phi_{\alpha}(p)$ (solid lines) of the Gaussian
potential energy operator in the momentum representation for oscillator length
$b=1$ fm and Bessel functions $j_{0}(pr_{\alpha})$ (dashed lines). Values of
$\alpha$ are indicated near the curves.}%
\label{Fig:V(p)_alpha_r0=1_gauss}%
\end{figure}

\begin{figure}[tbh]
\begin{center}
\includegraphics[
width=\textwidth]
{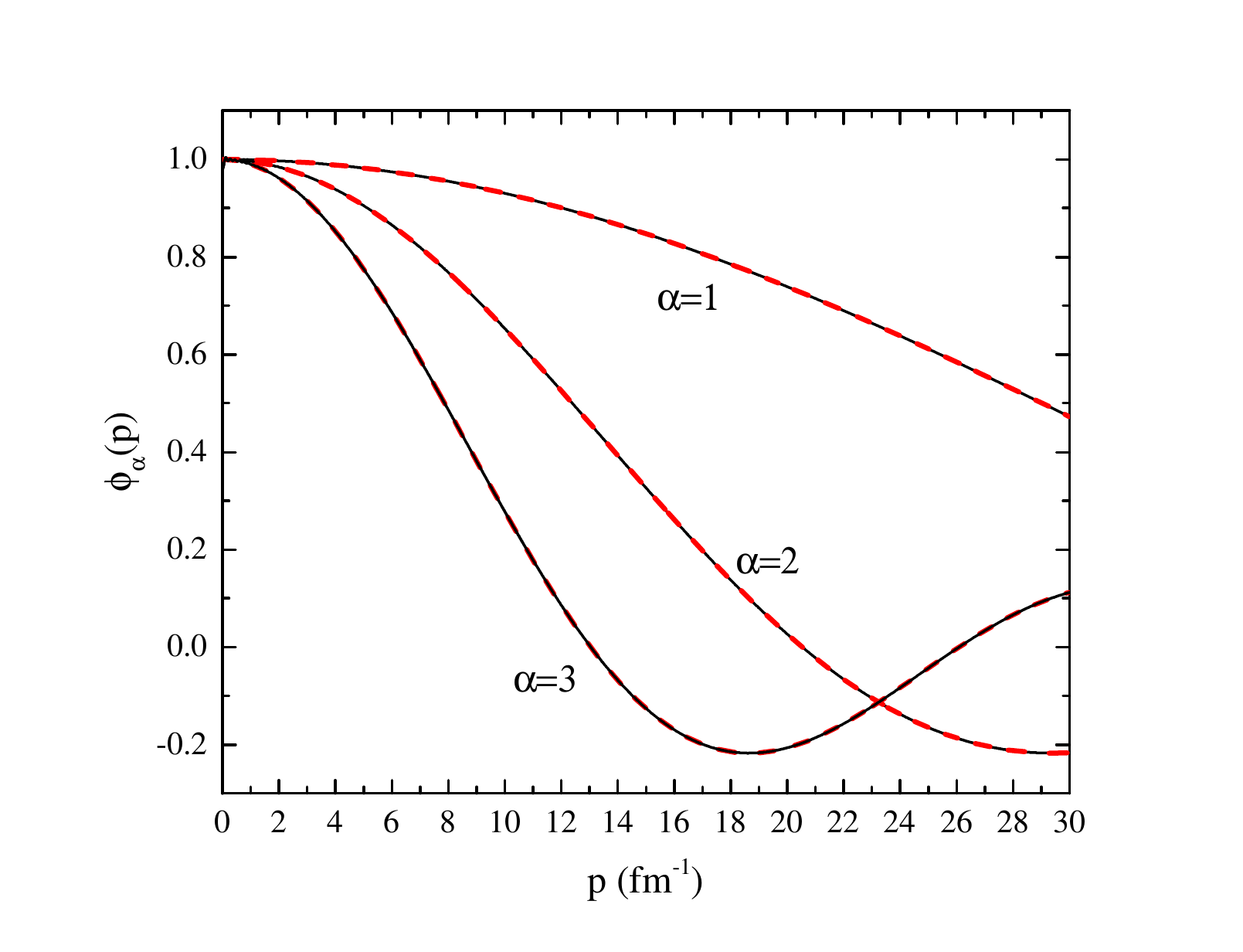}
\end{center}
\caption{Eigenfunctions $\phi_{\alpha}(p)$ (solid lines) of the Yukawa potential
energy operator in the momentum representation for oscillator length $b=1$ fm
and Bessel functions $j_{0}(pr_{\alpha})$ (dashed lines). Values of $\alpha$
are indicated near the curves.}%
\label{Fig:V(p)_alpha_r0=1_yukawa}%
\end{figure}

Fig. \ref{Fig:V(p)_alpha_r0=1_gauss} and Fig. \ref{Fig:V(p)_alpha_r0=1_yukawa}
display eigenfunctions $\phi_{\alpha}(p)$ of the Gaussian and Yukawa potential
energy operator in the momentum representation. The figures support our
conclusion that eigenfunctions $\phi_{\alpha}(p)$ are simply Bessel functions
$j_{0}(pr_{\alpha}^{N})$. Since the discrete coordinate $r_{\alpha}^{N}$
increases with $\alpha,$ the corresponding eigenfunction $\phi_{\alpha}(p)$
exhibits more oscillations with increase in $\alpha.$ Eigenfunctions of the
Gaussian and Yukawa potential energy operators are very similar, because
$r_{\alpha}^{N}$ depends only slightly on the location of the $\alpha$th zero
of the eigenfunction $U^{\alpha}_{n}$ in the discrete representation.

\subsubsection{Wave function of square-well potential.}

As one may notice, we do not show the eigenfunctions of the square-well
potential. They have a rather complicated form and may confuse the reader
of the present paper. The complexity originates from the degenerate states.
Indeed, in this case we have many degenerate eigenvalues 
 which coincide with the depth of the square-well
potential. Numerical diagonalization procedure reveals such eigenfunctions
which are a specific orthogonal combination of the correct (canonical)
eigenfunctions presented by Eq. (\ref{eq:SP167}). One may try to decompose
the nonstandard eigenfunctions in a set of standard ones. We do not perform
such a decomposition as it involves painstaking efforts and does not supply
us with new information.

In this respect it is important to notice that any potential, which contains
the repulsive and attractive parts, will or may have the twofold degenerate
eigenvalues, one of which corresponds to a repulsive part and the other
corresponds to an attractive part of the potential.  And thus their eigenfunctions
will be of a complicated unusual form. It is then necessary to study what
combinations of the standard eigenfunctions are presented in the obtained eigenfunctions.

\section{Conclusions \label{Sec:Conclusions}}

We have studied the main properties of the potential energy matrix for a model
problem - particle in a field of central-symmetric potential. We have
selected four types of potentials which are very often used both for model
and real physical problems. They are square-well, Gaussian, exponential and
Yukawa potentials. We have obtained and analyzed the structure of eigenvalues
and eigenfunctions of the potential energy matrix in a huge but finite basis
of oscillator functions. We demonstrated that eigenvalues of the matrix
coincide with the potential energy in some specific points of the coordinate
space, and eigenfunctions are the expansion coefficients of the spherical Bessel
functions over oscillator functions. These are the universal properties of
matrix elements of two-body potential matrix.

We have demonstrated that the large part of eigenvalues of the potentials
 equals zero. It means that only a very restricted number of the
potential eigenstates participates in the wave function of scattering state
and T-matrix.
The dependence of the eigenvalues on the
shape of potentials and oscillator length explains why and when we need a
small number of basis functions to obtain wave functions and scattering
parameters with the desired precision.

We have used an oscillator basis to construct matrix elements of the
potential energy operator and to study its eigenvalues and  eigenfunctions.
However, the eigenfunctions of the potential in the discrete representation will
be the expansion coefficients of the spherical Bessel function for any
square-integrable orthonormal basis, while the eigenvalues of the potential will
 be determined by a location of the zeros of these eigenfunctions.
The only difference will be in the explicit form of the eigenfunctions of the potential.

The results of the analysis of eigenvalues and eigenfunctions of the potential
energy matrix  performed in the present paper for model potentials will be
used to study properties of the potential energy operator for real physical
problems, namely, for a set of atomic nuclei which can be presented as a
two-cluster system. 

By comparing eigenvalues and eigenfunctions of the potential energy operator
obtained with and without (the so-called folding approximation) total
antisymmetrization, we are going to reveal effects of the Pauli principle in
nuclear two-cluster systems.

As we seen above, the diagonalization of the potential energy matrix proposes a self
consistent way of reducing a nonlocal potential (or nonlocal operator) to
the local form. It is intriguing to study what local form could suggest this
method for two-cluster systems. That will be a subject of the next paper.

After publication of the present paper on the site of electronic preprints, 
we were informed about the papers \cite{2001PhRvA..63f2708A, 
2007JPhB...40.4245N} where the approximate formula is used to calculate
matrix elements of different two-body potentials. This formula is similar to
the expression (\ref{eq:SP164}) we have deduced.  In
Refs. \cite{2001PhRvA..63f2708A, 2007JPhB...40.4245N} the approximate
method is based on the theorems for the numerical calculations of integrals
involving orthogonal polynomials. Such theorems can be found in the book \cite{Krylov_Integral}. To calculate matrix elements of the potential
energy operator, we have not used this approximate method. As pointed out above, we
have used the recurrence relations. Our results demonstrate that the
effective size of the matrix $\left\Vert U_{n}^{\alpha }\right\Vert $ in Eq.
(\ref{eq:SP164}) (i.e. the position of a remote node
determined in section (\ref{Sec:Extrapol})) depends
on the potential shape  and the oscillator length $b$. By comparing our
results with the approximate formula used in Refs. \cite{2001PhRvA..63f2708A, 
 2007JPhB...40.4245N}, we came to the conclusion that the approximate
method is very precise for relatively small values of the oscillator length
$b$ and for potentials without singularity.

\section*{Acknowledgements}
This work was supported in part by the Program of Fundamental Research of the
Physics and Astronomy Department of the National Academy of Sciences of
Ukraine (Project No. 0117U000239).

\end{document}